\documentclass[sigconf]{acmart}

\usepackage[linesnumbered,algoruled,boxed,lined]{algorithm2e}
\usepackage{algpseudocode}
\usepackage{makecell}
\usepackage{multirow} 
\usepackage{booktabs}
\usepackage{hyperref}
\hypersetup{
    colorlinks=true,
    linkcolor=blue,
    filecolor=magenta,
    urlcolor=magenta,
}
\usepackage[all]{hypcap}
\usepackage{booktabs}
\usepackage{url}
\usepackage{xcolor}
\usepackage{tikz}
\usepackage{tcolorbox}

\usepackage[framemethod=TikZ]{mdframed}
\definecolor{mycolor}{RGB}{194, 214, 236}
 
\newcommand{\tech}{\mbox{\textsc{AutoFirm}}}   

\newcounter{finding}
\newcommand{\finding}[1]{\refstepcounter{finding}
 	\vspace{1mm}
	\begin{mdframed}[linecolor=gray,roundcorner=12pt,backgroundcolor=gray!15,linewidth=3pt,innerleftmargin=10pt,innertopmargin=6pt,innerbottommargin=6pt,leftmargin=0cm,rightmargin=0cm,topline=false,bottomline=false,rightline = false]
		\textbf{Findings \arabic{finding}:} #1
	\end{mdframed}
	\vspace{0.5mm}
}

\newcounter{result}
\newcommand{\result}[1]{\refstepcounter{result}
 	\vspace{1mm}
	\begin{mdframed}[linecolor=gray,roundcorner=12pt,backgroundcolor=gray!15,linewidth=3pt,innerleftmargin=10pt,innertopmargin=6pt,innerbottommargin=6pt,leftmargin=0cm,rightmargin=0cm,topline=false,bottomline=false,rightline = false]
		\textbf{Result \arabic{result}:} #1
	\end{mdframed}
	\vspace{0.5mm}
}

\newtcolorbox{mybox}[2][]
  {colback = white, colframe = black,
    colbacktitle = gray, enhanced,
    attach boxed title to top left={yshift=-2mm,xshift = 4mm},
    title=#2,#1}

\makeatletter
\g@addto@macro{\@algocf@init}{\SetKwInOut{Parameter}{Parameters}}
\makeatother

\setcopyright{none}
\acmConference[Anonymous Submission to ACM XXX 2024]{}{}{TBD}
\acmYear{2023}

\begin{document}
\begin{sloppypar}

\title{{\tech}: Automatically Identifying Reused Libraries inside IoT Firmware at Large-Scale}

\author{YongLe Chen}
\affiliation{%
  \institution{Taiyuan University of Technology}
  \city{Shanxi}
  \country{China}
}
\email{chenyongle@tyut.edu.cn}

\author{Feng Ma}
\affiliation{%
  \institution{Taiyuan University of Technology}
  \city{Shanxi}
  \country{China}
}
\email{mafeng0473@link.tyut.edu.cn}

\author{Ying Zhang}
\affiliation{%
  \institution{Beijing JiaoTong University}
  \city{Beijing}
  \country{China}
}
\email{novzying@bjtu.edu.cn}

\author{YongZhong He}
\affiliation{%
  \institution{Beijing JiaoTong University}
  \city{Beijing}
  \country{China}
}
\email{yzhhe@bjtu.edu.cn}

\author{Haining Wang}
\affiliation{%
  \institution{Virginia Tech}
  \city{Virginia}
  \country{USA}
}
\email{hnw@vt.edu}

\author{Qiang Li}
\affiliation{%
  \institution{Beijing JiaoTong University}
  \city{Beijing}
  \country{China}
}
\email{liqiang@bjtu.edu.cn}

\begin{abstract}

The Internet of Things (IoT) has become indispensable to our daily lives and work. Unfortunately, developers often reuse software libraries in the IoT firmware, leading to a major security concern. If vulnerabilities or insecure versions of these libraries go unpatched, a massive number of IoT devices can be impacted. 
In this paper, we propose the {\tech}, an automated tool for detecting reused libraries in IoT firmware at a large scale.
Specifically, {\tech} leverages the syntax information (library name and version) to determine whether IoT firmware reuses the libraries. 
We conduct a large-scale empirical study of reused libraries of IoT firmware, investigating more than 6,900+ firmware and 2,700+ distinct vulnerabilities affecting 11,300+ vulnerable versions from 349 open-source software libraries. 
Leveraging this diverse information set, we conduct a qualitative assessment of vulnerable library versions to understand security gaps and the misplaced trust of libraries in IoT firmware. Our research reveals that: manufacturers neglected to update outdated libraries for IoT firmware in 67.3\% of cases; on average, outdated libraries persisted for over 1.34 years prior to remediation; vulnerabilities of software libraries have posed server threats to widespread IoT devices.

\end{abstract}

\maketitle

\section{Introduction}

In the Internet of Things (IoT) era, connected devices, ranging from surveillance cameras and routers to various home automation, automate daily routines and provide critical services to Internet users. 
Unlike conventional computer systems, IoT devices suffer from a larger attack surface~\cite{fernandes2016flowfence, he2018rethinking, jia2017contexlot, wang2018fear, fernandes2016security}, which includes weak credentials, insecure protocols, and insecure software. 
Successful exploration of IoT devices has posed a significant threat to critical infrastructure; for example, Mirai~\cite{antonakakis2017understanding} compromised hundreds of thousands of IoT devices via default credentials and launched several DDoS attacks against various online services like Krebs on Security and Dyn.

IoT firmware is the core software embedded in hardware devices, providing primary control, functions, and manipulation over device-specific peripherals. 
Toady's attack vectors of IoT devices usually originate from insecure/vulnerable firmware, enabling adversaries to access and take control of IoT devices.
One overlooked security issue is that the development of IoT firmware relies heavily on software libraries, which significantly improves development efficiency and lowers the cost.  
However, if a vulnerability is discovered in a reused library, this weakness could directly impact the IoT firmware.
Older/insecure versions of software libraries are often used in IoT firmware, which transitively affects many IoT devices.
If left unpatched, vulnerabilities or outdated versions in reused software libraries would pose direct or transitive security risks to IoT devices.

Securing IoT systems requires identifying vulnerable library usage in the firmware. 
Prior studies~\cite{camurati2018inception, redini2020karonte, xu2017neural, yu2020order, zhan2021atvhunter, kim2020firmae} have proposed static or dynamic methods to determine whether the vulnerable packages/codes are reused in firmware images, typically relying on the binary similarity detection between the original software and the library in firmware.
However, those approaches are not designed for the large-scale study of reused libraries in firmware due to three limitations. 
First, the firmware collection is a manual effort, which is time-consuming and labor-intensive.
Second, the firmware image is a binary file encapsulating the software, requiring manual efforts to convert the firmware into a list of reused libraries.
Third, the library version identification involves control-flow graph (CFG) extraction and binary similarity comparison.
If necessary, deep learning algorithms may be used to compare the two CFGs for similarity.

To address those limitations, we propose an automated tool, called {\tech}, to detect reused libraries in IoT firmware without manual effort. 
{\tech} consists of three components, including firmware collection, library identification, and vulnerable library detection.
Specifically, the firmware collection is to construct the dataset of IoT firmware images with their metadata description, e.g., device manufacturer names, product names, and release timestamps. 
Library identification consists of two stages: converting the firmware into a filesystem and extracting the list of reused libraries.
Vulnerable library detection leverages the syntax information (library name, version) to determine whether the IoT firmware reuses a vulnerable/outdated library version.
Overall, {\tech}'s inputs are the URL source of IoT firmware images, and the outputs are the tuple of library information, as (\textit{lib, version, CVE}). 
The core in the {\tech} is to automatically find vulnerable reused libraries.

Moreover, we conduct a large-scale empirical study of reused libraries inside IoT firmware, investigating 6,000+ firmware images for 2,700+ distinct vulnerabilities that affected a diverse set of 349 open-source libraries. 
We build our analysis on a dataset that merges IoT firmware from manufacturers' websites, vulnerability entries from the National Vulnerability Database~\cite{nvd}, and reused libraries in the firmware.
Specifically, we leverage (\textit{lib, version, CVE}) to combine libraries of IoT firmware and vulnerabilities.
Tying together those data sources would provide a large-scale analysis of software in IoT firmware, including an investigation of firmware characteristics and vulnerable library distribution.
We extensively analyze the firmware images containing vulnerable libraries to understand the underlying risks associated with IoT firmware.
Our research addressed three research questions: 
 
\noindent
\textit{$\bullet$ RQ1: When IoT firmware reuses an outdated library version, would the manufacturer update to a newer version?}

\noindent
\textit{$\bullet$ RQ2: How long does an outdated/vulnerable library version persist in IoT firmware?}

\noindent
\textit{$\bullet$ RQ3: What is the impact and influence of those vulnerable versions of libraries in IoT firmware? }

\textbf{Our Findings}. 
We have collected 6,901 IoT firmware images from 37 different IoT manufacturers' websites.
We find 11,342 vulnerabilities caused by 349 reused libraries over those 6,901 IoT firmware images, revealing that many vulnerable and out-of-date libraries are still used in IoT devices.  
There are 2,729 distinct CVEs among 11,342 vulnerabilities.  
Although IoT manufacturers should keep their devices updated and safe, this requirement is rarely met in practice. 
We provide a quantification study of IoT firmware~\cite{autofirm}, especially those that used software libraries and discovered vulnerabilities.
Among our findings, we identify that: in 32.7\% of the cases, the manufacturer will update a vulnerable software library to a newer version, compared with the persistence of the outdated version in 67.3\% of cases; the average time taken for the manufacturer to update the vulnerable library was approximately 1.34 years; vulnerabilities of libraries have posed server threats to widespread IoT devices.

In brief, our contributions are summarized as follows:
\begin{itemize}
  \item We develop and provide a open-source tool~\footnote{Artifact and Code: \url{https://github.com/sure17/AutoFirm}} to automatically detect reused libraries in firmware.
  \item We present the quantification analysis on IoT firmware, including firmware, reused libraries, and vulnerable library versions.  
  \item After analyzing 11,342 vulnerable versions (2,729 distinct CVEs) in 6,901 firmware images, we provide three findings for our research questions. 
\end{itemize}

\noindent\textbf{Roadmap}.
The remainder of this paper is organized as follows. 
Section 2 presents the design and implementation of {\tech}.  
Section 3 provides a large-scale analysis of IoT firmware and reused libraries. Section 4 provides discoveries and findings. 
Section 5 presents the discussion and Section 6 surveys the related work. Finally, Section 7 concludes.

\section{Background}

\begin{figure}[!t]
    \centering
    \includegraphics[width=2.4in]{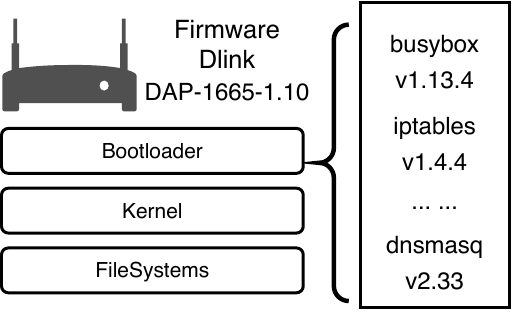}
    \caption{An example for software library usage in the IoT firmware.}
    \label{fig:back}
\end{figure}

IoT firmware is software that enables the device to perform functions necessary to make various hardware components work properly.
Figure~\ref{fig:back} depicts three components of the firmware (\textit{Dlink}, \textit{DAP-1665-1.10}), including the bootloader, the kernel, and the filesystem. 
The bootloader initiates the necessary hardware and system startup, and the kernel starts all processes required and additional services for the device to work. 
The filesystem stores all the individual files required for the device's performance, e.g., web servers and network services.
The filesystem of an IoT device can be of different types, depending on the manufacturer's requirements and the device's intended function. 
Software libraries with specific versions are located in the filesystem for IoT firmware. 
For instance, the busybox v1.13.4, iptables v1.4.4, and the dnsmasq v2.33 have been reused in the firmware (\textit{Dlink}, \textit{DAP-1665-1.10}), as shown in Figure~\ref{fig:back}.

Typically, software libraries in the IoT firmware have two issues: an outdated version and suffering a vulnerability. 
Developers heavily reuse open-source software libraries to implement the functions of IoT products. 
Firmware security relies on IoT manufacturers to keep their devices safe by updating the newest software library version to ensure their firmware withstands attacks. 
However, there exists a large deviation from the practice. 
Due to the lack of awareness, users are unaware of the vulnerabilities that may exist in an IoT ecosystem caused by the reused software libraries. 

\begin{figure}[!t]
    \begin{tabular}{ c   c  }
    \begin{minipage}[t]{0.46\linewidth}
    \centering
    \includegraphics[width=1.0\linewidth]{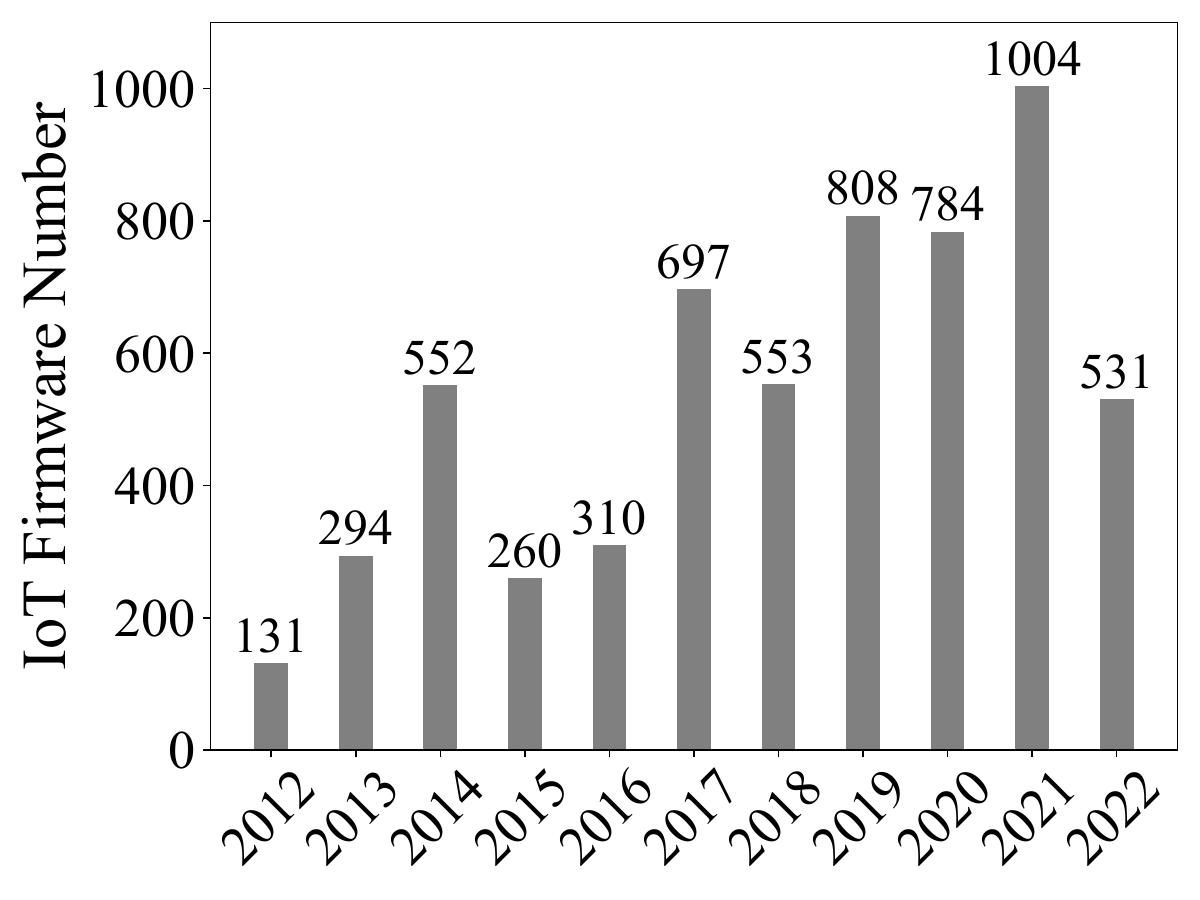}
    \caption{The number of IoT firmware along with time.}
    \label{fig:num:firmware}
    \end{minipage}
    &
    \begin{minipage}[t]{0.46\linewidth}
    \includegraphics[width=1.0\linewidth]{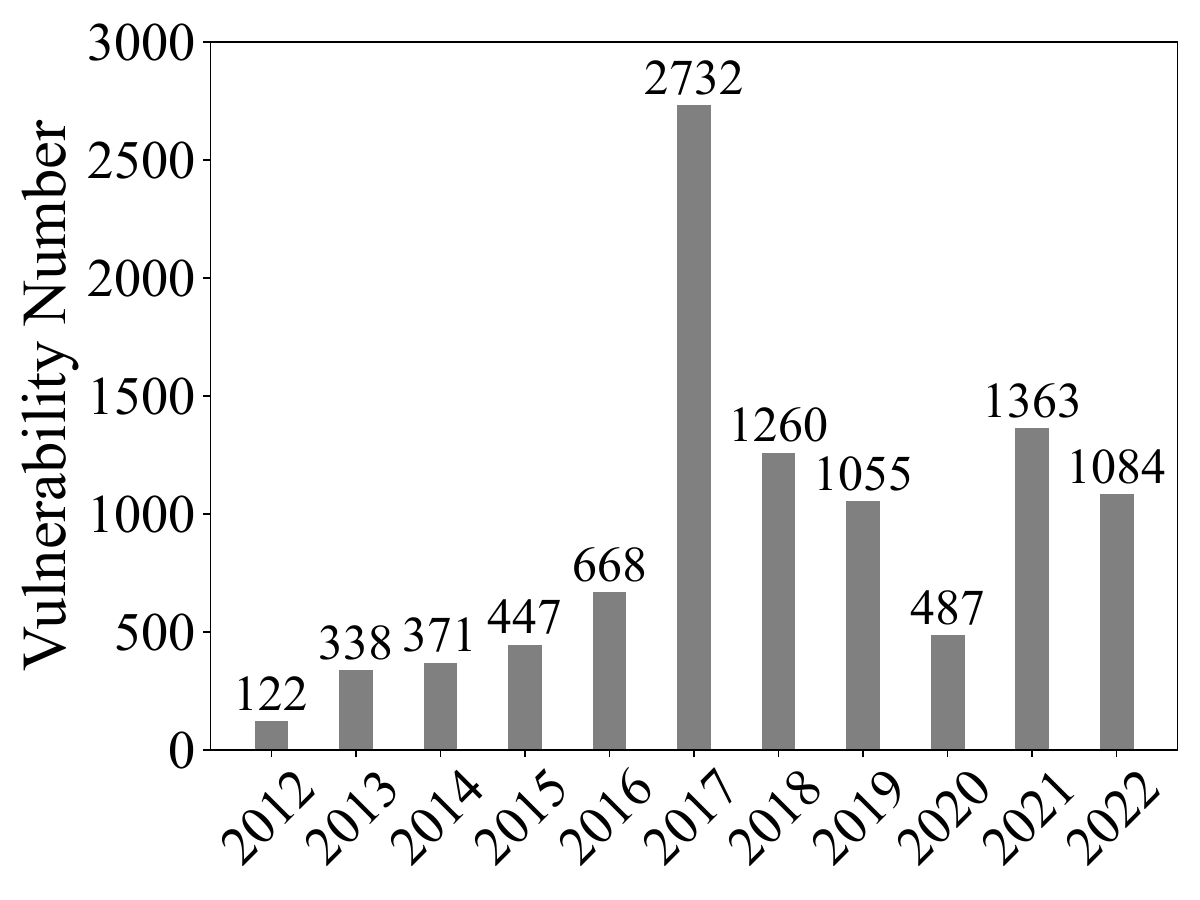}
    \caption{The vulnerability number of software libraries along with time.}
    \label{fig:num:lib}
    \end{minipage}
    \end{tabular}
\end{figure}

We provide a preliminary analysis of reused libraries in IoT firmware at a large scale. 
For the firmware number, we use the dataset collected by Firm2Lib to illustrate the IoT firmware in the market.
Figure~\ref{fig:num:firmware} depicts the increasing number of IoT firmware along with time, where the X-axis is the timeline by year, and the Y-axis is the number of IoT products released on their manufacturer websites.
It is evident that the number of IoT device firmware is growing.
For the library number, we use the national vulnerability database (NVD)\cite{nvd} to demonstrate the scale of the software.
If the vulnerability involves the software library, we extract this CVE.
Figure~\ref{fig:num:lib} depicts the vulnerability number of software libraries along with time, totally 
6,938 CVEs. 
Similarly, we find that more and more software libraries are suffering vulnerabilities with time.
Combining Figure~\ref{fig:num:firmware} and Figure~\ref{fig:num:lib}, we need a large-scale detection for discovering the library in IoT firmware. 
However, the large-scale study of reused libraries in firmware suffers three limitations: (1) the firmware collection is a manual effort; (2) converting the firmware into a list of reused libraries is a manual process; and (3) identifying reused library version requires binary similarity approach.

\section{Design and Implementation}

In this section, we present the design of {\tech}, as an automated tool for detecting the vulnerable libraries of IoT firmware at a large scale.

\textbf{Our Motivation}. 
Today's software packages usually write their syntax information in the header of their binary files. 
The syntax information in the hexadecimal format may contain several fields related to the software libraries: authors, released time, version, copyright, and license. 
If the library's syntax (name and version) matches the vulnerability information, we can detect whether it is reused in the IoT firmware. 
We leverage the library syntax to identify the firmware's library suitable for large-scale detection. 
Our tool has two advantages for the large-scale detection of the firmware's library: real-time and automatic detection. 
In other words, our tool can detect reused libraries of IoT firmware in a short period without manual effort.

\begin{figure}[!t]
    \centering
    \includegraphics[width=3.3in]{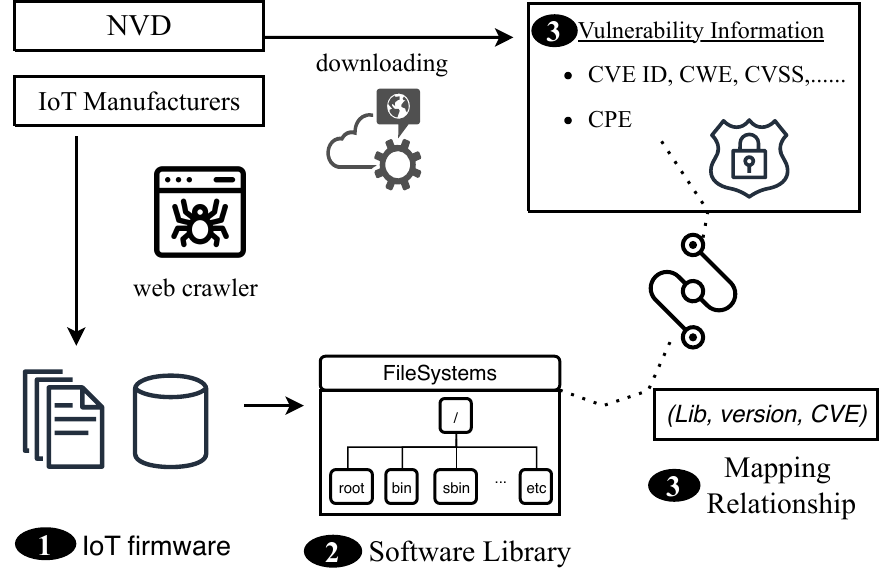}
    \caption{An overview of {\tech}'s architecture: (1) IoT firmware collection, (2) Library list identification, and (3) Vulnerable library detection. }
    \label{fig:arch}
\end{figure}

\textbf{Architecture}.
Figure~\ref{fig:arch} depicts an overview of {\tech}'s architecture.
{\tech} centered around IoT manufacturers' websites, the National Vulnerability Database (NVD)~\cite{nvd}, and affected libraries from IoT firmware.
In particular, we use the web crawler to download IoT firmware images, which are subsequently converted into filesystems to locate affected software libraries.
The filesystem comprises several directories, such as ``bin'' or ``sbin'' and the corresponding binary files.
{\tech} searches libraries in the filesystem of IoT firmware to identify the library name and looks for strings (version) in the software's hexadecimal file. 
The vulnerability information specifies if a vulnerability or a patch is available for a specific library version in IoT firmware.
By matching the library name and version with the vulnerability information, we can establish a mapping relationship between the firmware library and the vulnerability.
Below, we present the details of three components in {\tech}.

\subsection{IoT Firmware Collection} 

An available dataset is necessary and preliminary for the large-scale analysis of libraries of IoT firmware images. 
Typically, downloading firmware on the Internet is the easiest and most common way. 
To the best of our knowledge, there are no public datasets for IoT firmware.
Thus, we have undertaken extensive research to identify dozens of popular IoT manufacturers (detailed in Table~\ref{tab:download:list} in the Appendix), like D-Link and TP-Link, which offer download URLs for their firmware. 
These firmware images are usually found on their official website's device support or download pages.
Additionally, we have discovered that discussion forums and GitHub repositories may also provide links to firmware that developers or security professionals have voluntarily collected. 
We have deployed the web crawler to periodically collect IoT firmware images from the Internet to construct a large-scale raw dataset.
The module's input is the URL of the IoT manufacturer's website, and the output is the dataset of IoT firmware images.

Specifically, we have leveraged the Scrapy framework~\cite{scrapy} to implement the web crawler. 
There are several practical issues arise when collecting IoT firmware images.
Firstly, today's websites have imposed limitations on the web crawler, including asynchronous loading, javascript encryption, and dynamic cookies.
Hence, we eliminate those limitations via the BrowserMob Proxy~\cite{browsermob}, where we manipulate all HTTP requests and responses to capture HTTP content and export firmware images as files.  
Secondly, IoT firmware should contain metadata that describes relevant information such as device type (e.g., router or webcam), manufacturer (e.g., D-Link or TP-Link), and product model (e.g., TD-8840T).
We extracted this information and stored it in JSON format. 
We remove the firmware image from the dataset if the metadata is missing. 
Lastly, different websites may provide duplicate firmware images, and to avoid this, we use the checksum of firmware metadata to filter out duplicates in the web crawler.
One example is shown in Table~\ref{tab:mata} in the Appendix, including firmware name, manufacture name, device type, product name, version, publish time, URL, and checksum.

\subsection{Library Identification}

We extract libraries from IoT firmware images.  
These libraries are often embedded and executable binary files within the firmware's filesystem.
The library identification contains two stages: (1) extracting a filesystem from a firmware image and (2) finding reused libraries in the filesystem.

Every firmware image is converted into a folder as its filesystem.
Specifically, we have utilized binwalk~\cite{binwalk} to extract the filesystem of IoT firmware.
The binwalk is an open-source tool for analyzing, reverse engineering, and extracting firmware. 
However, some IoT firmware may be compressed, which can cause practical issues when finding libraries. 
In these cases, we use standard compression algorithms, such as zip or tar, to unarchive the firmware images. Once we have converted the firmware image into a folder, we iteratively decompress any remaining files until no compressed files are found. 
Note that any unknown compressed data is skipped.
Additionally, firmware may be encrypted, which can make it difficult to find its reused libraries. In these cases, we filter out the encrypted firmware based on the entropy of the firmware binary code. If the entropy in different offsets is stable, we consider the firmware encrypted.

Once the IoT firmware is unarchived, we explore all files within its filesystem to find libraries.
We have utilized the library names to find the libraries in the filesystem. 
Table~\ref{tab:software} lists four categories of 349 libraries used in IoT firmware: cmd tools, built-in components, UNIX tools, and open-source software. The `cmd' category includes commands controlling the device, such as `ls' and `ps'.
The `Built-in'' category indicates that libraries are included with the device and cannot be removed, such as `upagent'.
The `UNIX tool'' category indicates a common tool used in UNIX-based systems, such as `binutils'.
The `Open-source'' indicates that libraries are reused from the open-source community, such as `busybox'.

We find a reused library if a file name matches a library name.
One practical issue here is that file names in the filesystem differ from official software names.
When the software libraries run on IoT firmware, their names may be aliases or abbreviations rather than the original names.
To address this issue, we store all possible file names for each reused library in Table~\ref{tab:software}, where those terms are collected offline.
The collection approach is heuristic as follows. (1) We inspect the syntax information in the hexadecimal header of the library file. (2) If the syntax information contains software names, authors, and sources (URL), we extract those field values. 
(3) We use the Google search engine to determine if the file is a software library. If we can find the library's official website or GitHub repository, its file name is stored as a candidate term.

\begin{table}[!t]  \small
	\caption{Reused library terms. }
	\label{tab:software}
	\centering
	\begin{tabular}{c c}
		\toprule
		Category  &	   Num. \# (Terms) \\
		\midrule
		CMD & 24 (file)\\
		Built-in components & 85 (upagent)\\
		UNIX Tools & 21 (binutils)\\
		Open-source libraries & 219 (busybox)\\
		\bottomrule
	\end{tabular}
\end{table}

\subsection{Vulnerable Library Detection}

Vulnerable library detection is to determine whether a software library in IoT firmware is vulnerable, including two stages: (1) vulnerability information extraction, and (2) matching libraries with vulnerabilities.

Researchers and vendors heavily rely on the National Vulnerability Database (NVD)~\cite{nvd}, a repository that collates publicly disclosed vulnerabilities and assigns each a unique Common Vulnerabilities and Exposures (CVE) identifier. 
Whenever an IoT vulnerability is discovered, security researchers or vendors can request a CVE ID from the CVE Numbering Authority, which is subsequently associated with the identified vulnerability. 
Initially, this information may not be publicly disclosed, but as soon as it becomes available, it is added to the NVD's CVE list.
Upon public disclosure, the NVD adds additional data, such as vulnerability summaries, links to external references like security advisories and reports, the Common Vulnerability Scoring System (CVSS)~\cite{cvss}, enumeration of the affected software, and classification under the Common Weakness Enumeration (CWE)~\cite{cwe}.
We downloaded the NVD XML dataset and integrated it into {\tech}, as snapshotted in 2023, including 110,238 CVE vulnerabilities.

We combine reused libraries and vulnerabilities via string matching. 
Every library has the syntax information as the form (name, version).
A vulnerability has a common platform enumeration (CPE) describing software-relevant information.
For instance, the library ``samba`` has a CPE string ``cpe:/a:samba:samba:4.0'', where ``a'' indicates the application, ``samba:samba'' indicates the software name and its owner organization name, and ``4.0'' is the version.  
To build their mapping relationship, we compare the (name, version) with the CPE string of known vulnerabilities.
Regarding the library syntax, we use a rule to determine a vulnerable library: if its name and version match the CVE information, the vulnerable library is reused in the IoT firmware.

We have used the String library and the QEMU~\cite{qemu} tool to extract the printable strings of the header of reused libraries. 
Specifically, we adopted two manners of extracting the version information from the binary library. 
(1) We extract a list of null-terminated strings of printable characters from the hexadecimal header. 
The version is usually located in the first few bytes of the file.  
(2) We use the QEMU~\cite{qemu} tool to emulate the software library and output the relevant information. 
For the library version, we use the standard version scheme (``X.Y.Z'') and the comparison operators (``''$\geq$'', ``$\leq$'', ``$\sim$'', and ``''$\land$'').

\begin{table}[!t] \small
	\caption{ Distinct name set of the regex for libraries in IoT firmware.}
	\label{tab:software:regex}
	\centering
	\begin{tabular}{c}
		\toprule
		Version Regex \\
		\hline 
		\makecell[l]{[component][a-z$\backslash$s$\backslash$-$\backslash$\_$\backslash$.]*}  \\ \hline
		\makecell[l]{(0\textbar[1-9]$\backslash$d\{0,3\})(?:$\backslash$.$\backslash$d\{1,2\})\{1,2\}([a-z])?\\(?:-((?:0\textbar[1-9]$\backslash$d*\textbar$\backslash$d*[a-zA-Z-][0-9a-zA-Z-]*)\\(?:$\backslash$.(?:0\textbar[1-9]$\backslash$d*\textbar$\backslash$d*[a-zA-Z-][0-9a-zA-Z-]*))*))? } \\
		\bottomrule
	\end{tabular}
\end{table}

\begin{table*}[!t] \small
    \caption{ The IoT manufacturers with corresponding firmware number and success rate for extracting filesystems. }
    \label{tab:extract}
    \centering
    \begin{tabular}{c c c c c c }
    \toprule
    Manufacturer & \#  Firmware Num.  & \# Extracted Num.   & Manufacturer & \#  Firmware Num.  & \# Extracted Num.  \\
    \toprule
    360      &      20              & 20(100\%)  &  Xerox &         146             & 119(82\%) \\
    ASUS    &       322             & 332(100\%)  & Ubiquiti &      1,156           & 1,100(95\%) \\
    Buffalo &           8            & 8(100\%) & Trendnet  &     431             & 417(97\%) \\
    DLink &         557             & 534(96\%)  & TP-Link  &      1,978            & 1,818(92\%)  \\
    Foscam &        102              & 100(98\%)  & Tomato-Shibby & 321             & 313(98\%)  \\
    Linksys &       27              & 27(100\%)  & TI &            11              & 7(64\%)  \\
    Mercury &       324             & 318(98\%)  & Tenda &         318             & 318(100\%)  \\
    Microstrain &   3               & 3(100\%)   &   Synology &      61              & 61(100\%)  \\
    Mikrotik &      55              & 55(100\%)  & Supermicro &    439             & 439(100\%)  \\
    Netgear &       245             & 229(93\%)  & SE &            32              & 29(91\%)  \\
    Qnap &          335             & 318(98\%)  \\
    \toprule
    \end{tabular}
\end{table*}

One practical issue is that strings in the binary file's header contain much other information besides the library version. 
In this case, identifying semantic elements (Named Entity Recognition~\cite{zhou2002named} and Relation Extraction~\cite{zelenko2003kernel}) are typical techniques for extracting version elements from the strings, which have been extensively studied in the Natural Language Processing (NLP) community. 
However, we use state-of-the-art tools like Stanford NER, which leads to low precision and recall for extracting version elements from the strings. 
The reason is that existing NLP techniques cannot be directly applied to version-related context discovery.
We write the regex (Table~\ref{tab:software:regex}) to recognize version elements, which works well in practice.

\subsection{Implementation}

We have implemented the prototype of {\tech} as the tool~\cite{autofirm} for the large-scale detection of vulnerable libraries in IoT firmware. 
For the firmware collection module, we have leveraged the Scrapy framework~\cite{scrapy} to implement web crawler scripts for all IoT manufacturers, where the BrowserMob Proxy~\cite{browsermob} manipulates dynamic webpages of manufacturers' websites.
Our crawler scripts would directly download firmware images and store them as binary files with their metadata in JSON format.
For the library identification module, we have utilized the binwalk~\cite{binwalk} to extract the filesystem of IoT firmware and the regex to find the candidate set of reused libraries. 
Every firmware image is converted into a folder as its filesystem, and its reused libraries are stored as the name and located path in JSON format. 
For the vulnerable library detection module, we have used the String library and the QEMU~\cite{qemu} tool to extract the printable strings in the header of libraries. 
We store syntax information for every library as the form (library name, version) and compare it with the CPE string of known vulnerabilities from the NVD~\cite{nvd}.
{\tech} connects those 3 modules via the pipeline script, where the input is the website URL, and the output is the vulnerable libraries in IoT firmware.
\section{Large-scale Analysis}

In this section, we have systematically analyzed various characteristics of firmware libraries at a large scale, including firmware, resued libraries, and vulnerable libraries.

\subsection{Firmware Characteristics}

Table~\ref{tab:extract} lists the number of 21 different IoT manufacturers. IoT firmware covers popular device types, such as routers, switches, cameras, and printers.
In total, we have collected 6,901 firmware images from 21 different IoT manufacturers. 
The majority of these images (93\%) have corresponding metadata, including device types, manufacturer names, supportable product models, and timestamps. 
Nearly 7\% of IoT firmware does not have metadata information. 
From the vendor's perspective, TP-Link has the largest number of firmware images (1,978), while Ubiquiti has the second-largest number (1,156). 
Note that these firmware images use different file types to represent their binary formats, where {\tech} converts the binary file into the filesystem.  
The ``bin'' format has the largest number of files, almost 3,944, and other formats (e.g., ``img'', ``tar'', or ``zip'') have the second largest number. 
One practical issue is that some firmware images use an unsupportable format we cannot analyze further.
For instance, the manufacturer ``Supermicro'' usually uses numbers (e.g., ``.605'') as the file format of firmware images.
We utilize two heuristic rules to filter out the unqualified firmware: (1) removing a firmware image when its suffix format length is more significant than 5; (2) removing a firmware image when its suffix format contains more than one number.

\begin{figure}[!t]
    \centering
    \begin{tabular}{ c   c  }
        \begin{minipage}[t]{0.47\linewidth}
            \includegraphics[width=1.0\linewidth]{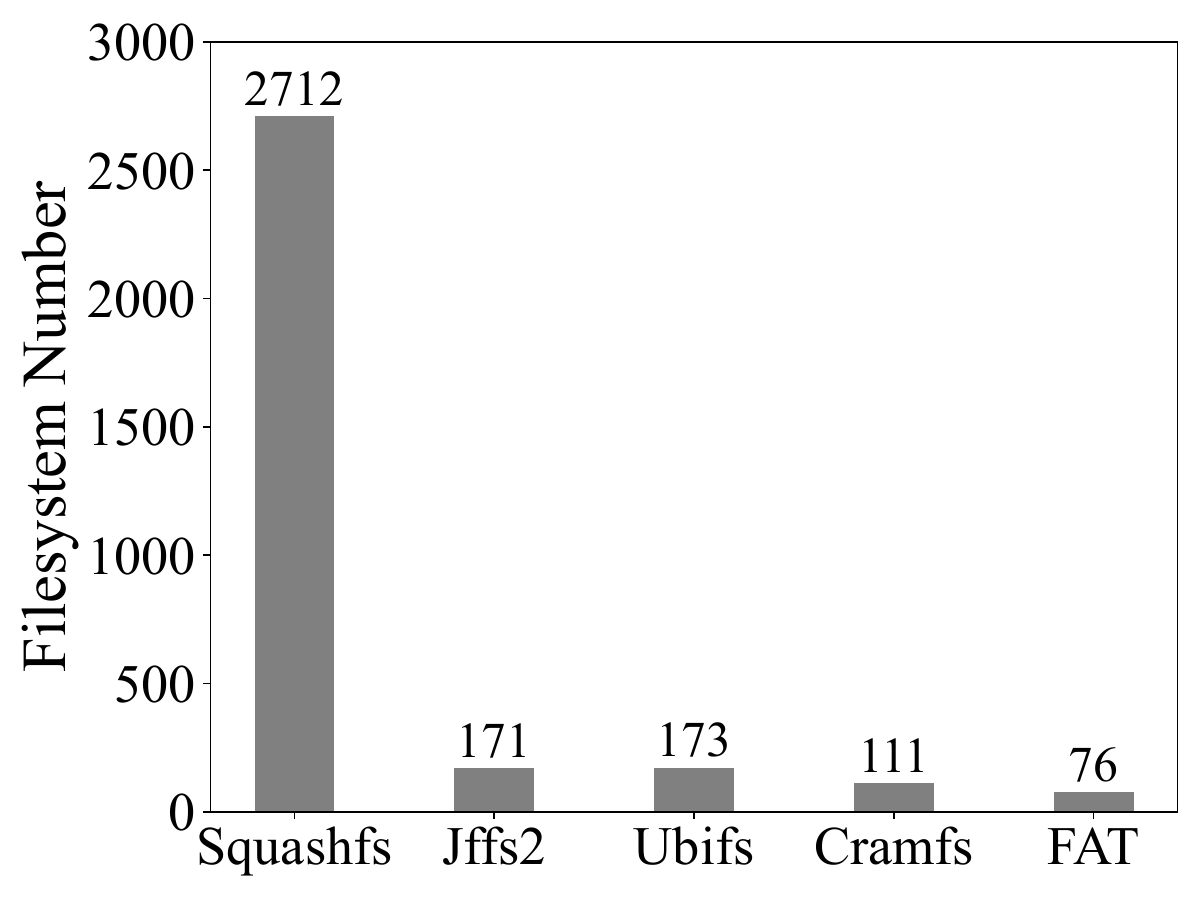}
            \caption{ The distribution of firmware filesystem.}
            \label{fig:filesystem}
            \end{minipage}
    &
    \begin{minipage}[t]{0.47\linewidth}
        \includegraphics[width=1.0\linewidth]{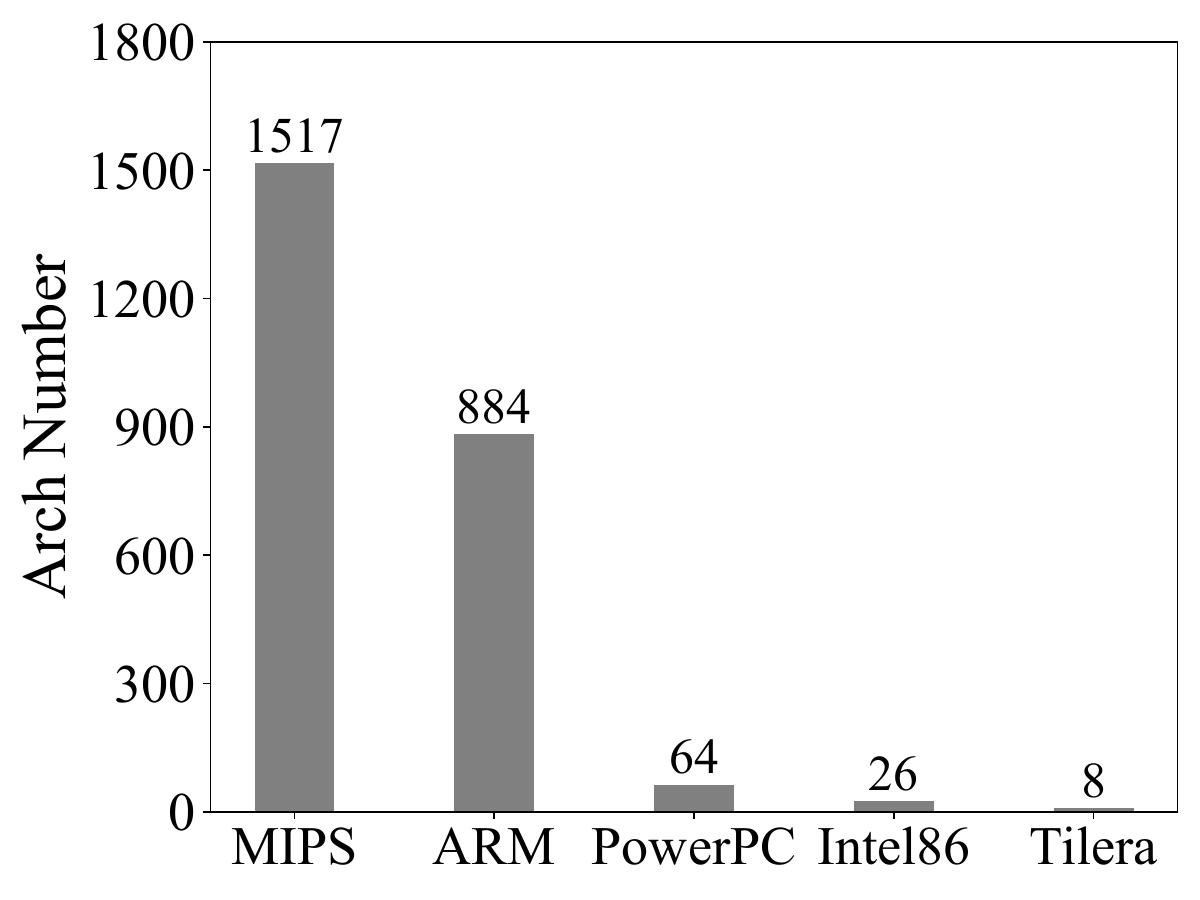}
        \caption{ The distribution of firmware architecture.}
        \label{fig:exp:arch}
        \end{minipage} 
    \end{tabular}
\end{figure}

We have provided an analysis of firmware characteristics, including the filesystem, architecture, and operating system (OS). 
We use the hexadecimal offset of the binary file header to determine those firmware characteristics. 
Figure~\ref{fig:filesystem} lists the distribution of the firmware filesystem, which is also diverse, with `Squashfs', `JFFS2', `Yaffs2', and `Ext2' being the most common. 
It is obvious that `Squashfs' covers 2,712 IoT firmware, which is a compressed read-only file system for Linux. 
Note that `Squashfs' compresses all files of firmware, indicating {\tech} has to decompress the IoT firmware.  
The distribution of the firmware architecture is shown in Figure~\ref{fig:exp:arch}, with MIPS and ARM being the two most common architectures for IoT devices. 
The reason is straightforward that MIPS and ARM have an advantage in low power consumption, making them as a popular choice for mobile and embedded devices.
We have identified the OS types of IoT firmware images, with Linux being the most common, followed by VxWorks, Cisco OS, WindowsCE, and Minix.

\begin{figure}[!t]
    \centering
    \includegraphics[width=2.5in]{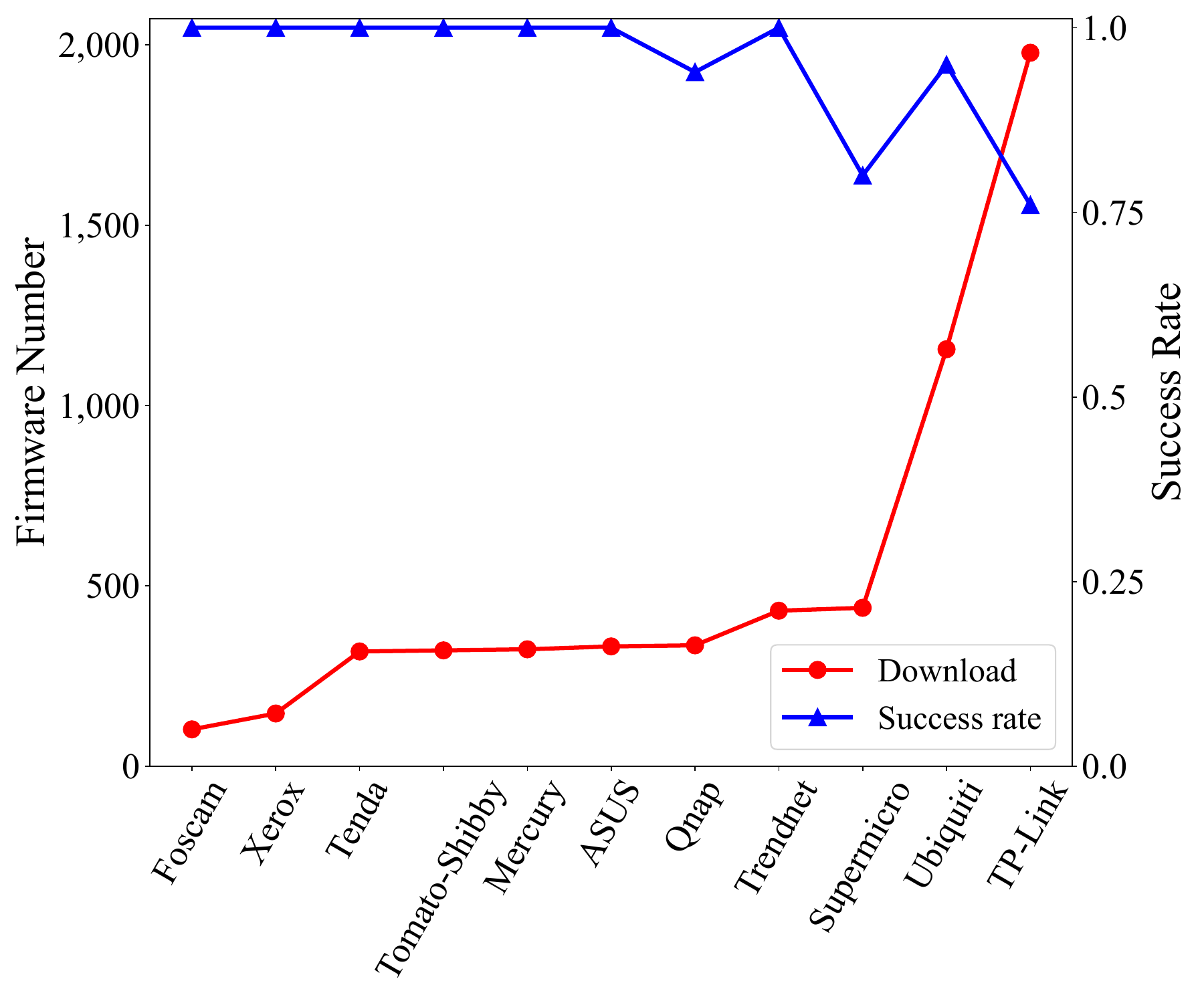}
    \caption{ The distribution of downloading firmware images from the Internet: the red-color curve represents the firmware number, and the blue-color curve indicates the success rate.}
    \label{fig:download}
\end{figure}

\colorbox{mycolor}{\textit{Download Analysis.}}
We have investigated 37 IoT manufacturer websites, yet, only 21 manufacturer websites (Table~\ref{tab:extract}) still provide a web service for firmware collection. 
This means that many manufacturers do not provide any web service for firmware downloads or have discontinued their services.
This lack of available web services (16 out of 37) is the major obstacle to conducting a large-scale empirical study on IoT firmware.
Some websites explicitly block web crawlers from accessing their firmware resources, while others make firmware images unavailable, likely due to server maintenance, updates, or other reasons. 
We evaluate the download success rate from 10 IoT manufacturer websites to demonstrate the firmware collection, as shown in Figure~\ref{fig:download}.
We observe that the success rate of firmware collection varied significantly, ranging from 40\% to 100\%. This indicates that the availability of firmware images on the Internet depends on individual manufacturers and their web services.
\result{
While downloading firmware from the Internet remains a feasible option for large-scale vulnerability detection, changes in the availability of firmware resources over time pose significant challenges. 
}

\subsection{Reused Library Characteristics}

\begin{table}[!t] \small
    \caption{ Top-5 Vendors: the total number of libraries over 4 categories. }
    \label{tab:dis:lib}
    \centering
    \begin{tabular}{c c c c c }
    \toprule
    Manufacturer & CMD  & Built-in & UNIX Tools &  Open-source  \\
    \toprule
    Tomato-Shibby   & 879   & 380   & 384       & 5,826 \\
    TP-Link         & 701    & 578    & 579       & 7,592  \\
    Trendnet        & 305   & 122   & 123       & 1,380       \\
    Ubiquiti        & 976   & 365      & 370       & 5,926       \\
    ASUS         &      266    & 294        & 261        & 2,159 \\ 
     \toprule
    \end{tabular}
\end{table}

\begin{figure}[!t]
    \centering
    \includegraphics[width=2.5in]{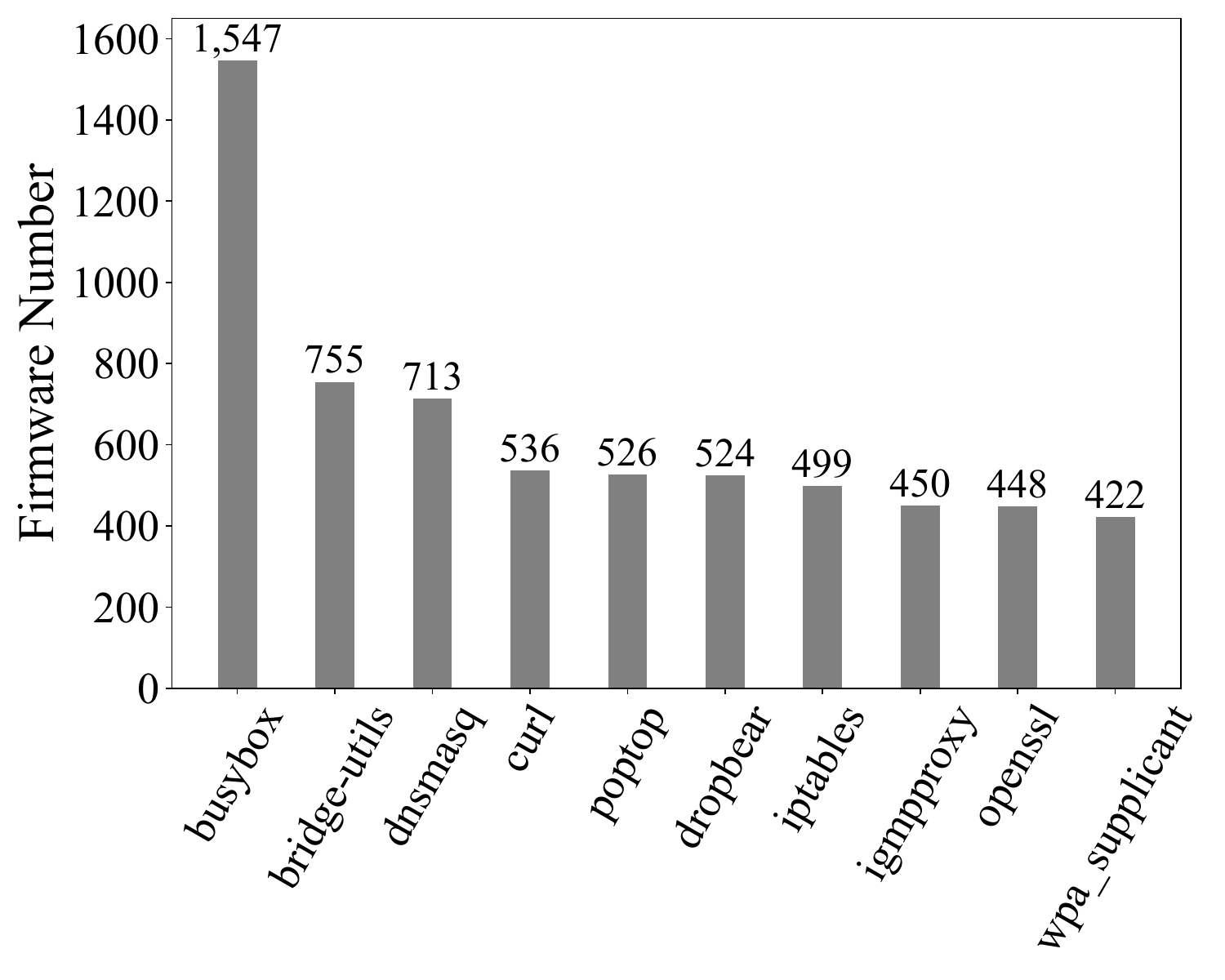}
    \caption{ The top-10 popular reused libraries reused among IoT firmware images.}
    \label{fig:lib:across}
\end{figure}

We have converted 6,901 firmware images into 6,582 filesystems to find the reused library.
Overall, the filesystem extraction success rate of 95\% by the binwalk tool~\cite{binwalk}.
We have provided 4 categories to those reused libraries based on their usage: ``cmd'', ``built-in'', ``UNIX tool'', and ``Open-source''. 
Table~\ref{tab:dis:lib} lists the top five manufacturers (`Tomato-Shibby', `TP-Link', `Trendnet', `Ubiquiti', and `ASUS') whose filesystems consist of the most reused libraries.
Our analysis shows that `Tomato-Shibby' has the largest number of extracted reused libraries, followed by `TP-Link', `Trendnet', `Ubiquiti', and `ASUS'.  
This discovery suggests that these vendors may use more software libraries in IoT firmware, potentially making their devices more vulnerable to software vulnerabilities.
As shown in Figure~\ref{fig:lib:across}, the top 10 most commonly reused libraries in IoT firmware images are presented. 
Note that the `busybox' and the `bridge-utils' are the most widely reused libraries, appearing in 40.3\% and 14.4\% of the firmware images, respectively.
The high reuse of specific libraries across IoT firmware images highlights the importance of identifying and patching vulnerabilities in these libraries.

\begin{figure}[!t]
    \centering
    \begin{tabular}{ c   c  }
    \begin{minipage}[t]{0.46\linewidth}
        \includegraphics[width=1.0\linewidth]{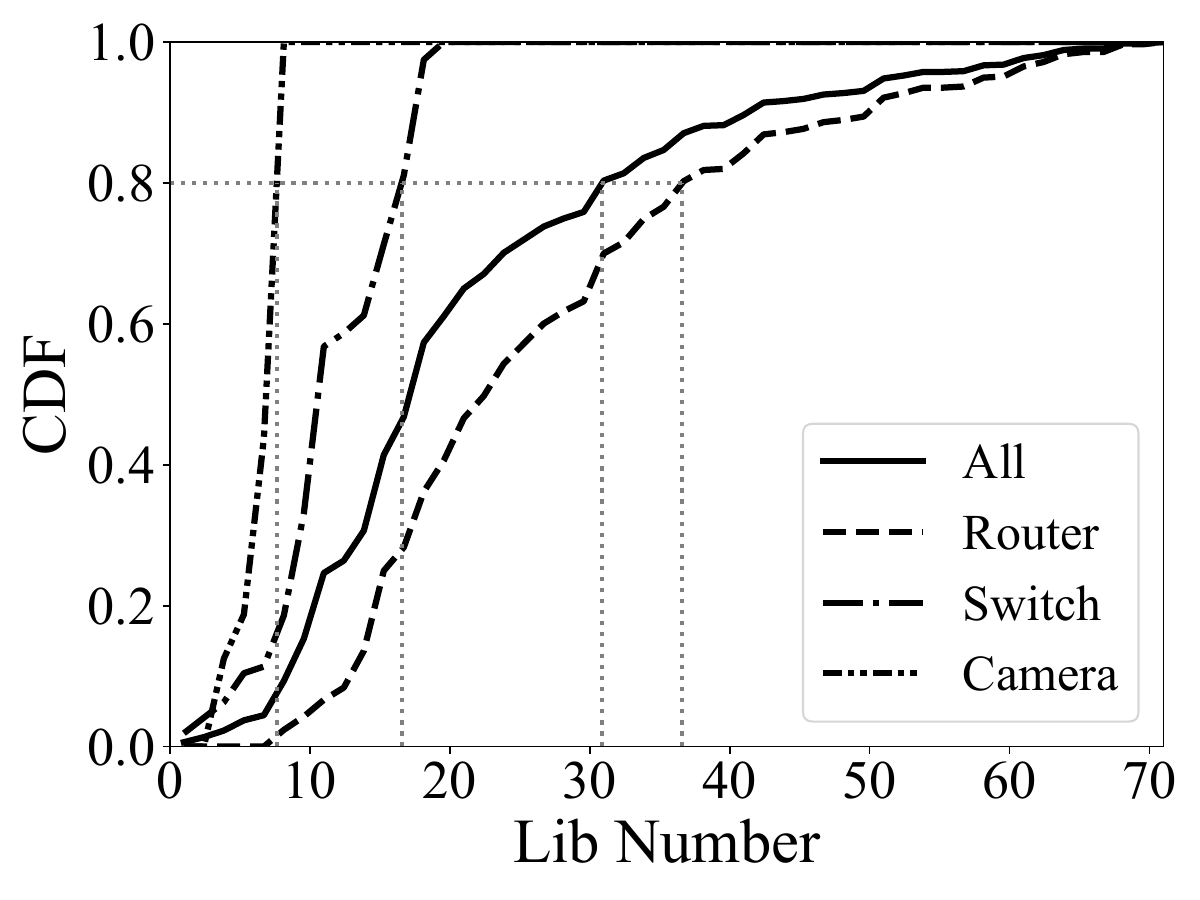}
        \caption{ Device type: The CDF of reused libraries per IoT firmware.}
        \label{fig:sle:type}
    \end{minipage}
    &
     \begin{minipage}[t]{0.46\linewidth}
            \includegraphics[width=1.0\linewidth]{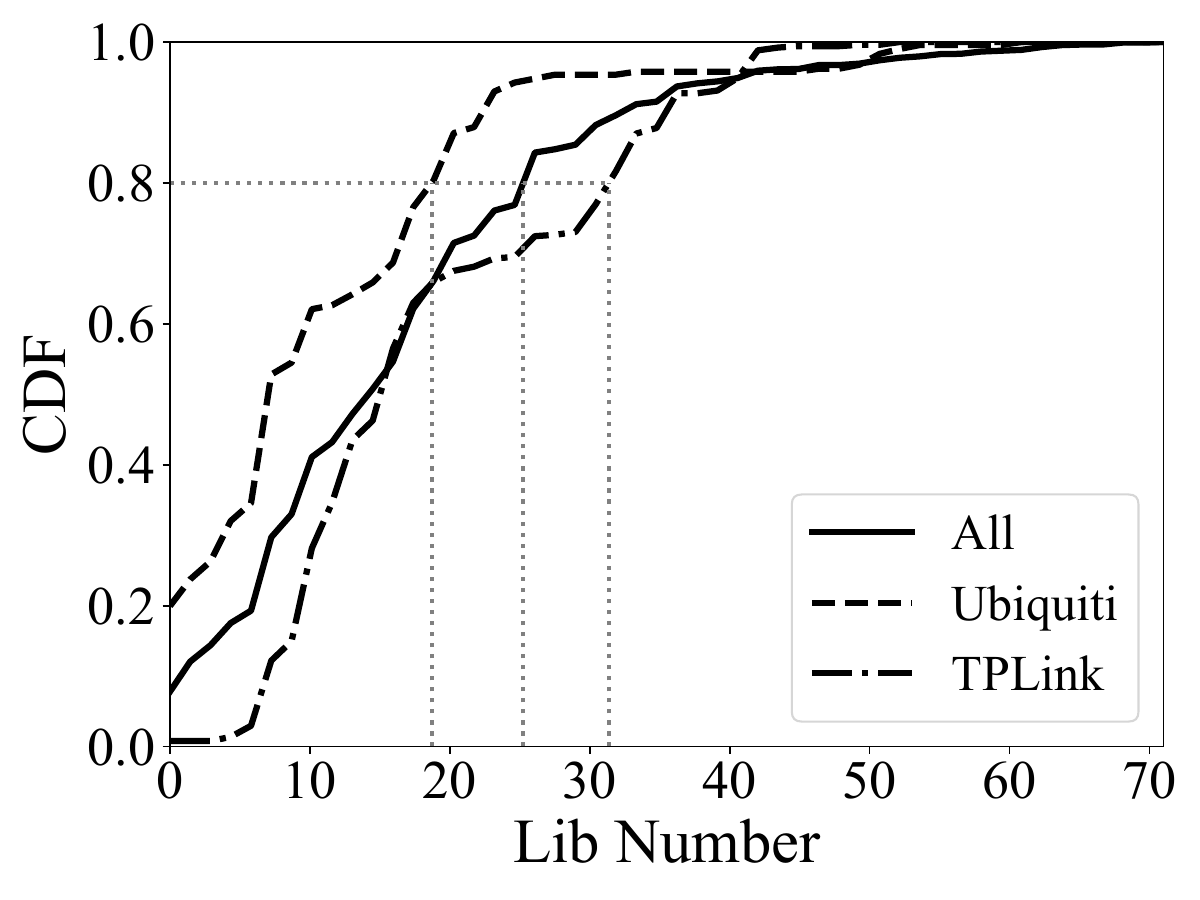}
            \caption{ Vendor: The CDF of reused libraries per IoT firmware.}
            \label{fig:sle:vendor}
        \end{minipage}
    \end{tabular}
\end{figure}

\colorbox{mycolor}{\textit{Usage Analysis.}}
From the perspective of the device type, Figure~\ref{fig:sle:type} depicts the distribution of the number of libraries per firmware, including the router, the switch, and the camera. 
It is intriguing to note that 80\% of routers contain 36 or fewer reused libraries, switches have fewer than 17 libraries, and cameras have eight or fewer reused libraries.
The number of reused libraries in IoT firmware can provide insights into the complexity and functionality of the device.
Interestingly, there is a significant gap between the number of libraries in camera devices and routers, suggesting that camera devices have a more specialized and complex software stack than routers.
From the perspective of the device vendor,  Figure~\ref{fig:sle:vendor} depicts the distribution of the number of libraries per firmware.
The distribution of the number of libraries shows that 80\% of IoT firmware contains 25 or fewer libraries (the black-color line), and different IoT vendors (Ubiquiti and TP-Link) have a similar distribution.
This distribution is consistent across different vendors, such as Ubiquiti and TP-Link, suggesting a typical usage pattern in IoT firmware development.
In general, a larger number of libraries can imply a more complex device with a wider range of functionalities.

\result{
The number of reused libraries is related to the functions of IoT devices. Different types of devices have different numbers of reused libraries, and the reused libraries of different vendors are similar.
}

\begin{table}[!t] \small
    \caption{The success rate of extracting version information from top 10 libraries.}
    \label{tab:vld:rate}
    \centering
    \begin{tabular}{ c c c | c c c}
    \toprule
    Library   &  Version   & Suc rate  &   Library   &  Version  & Suc rate\\
    \toprule
    busybox  & 2,117     &  80\% & dnsmasq & 773 & 95\%  \\
    mtd-utils & 3,055    & 97\% & radvd & 413    &    85\%  \\
    wpa\_supplicant & 680 & 92\% & ntfs-3g & 517 & 93\%  \\
    lighttpd & 124      & 86\% & iptables & 891  & 84\%  \\
    hostapd & 667       &  92\% & dropbear & 608 & 88\%  \\
     \toprule
    \end{tabular}
\end{table}

\subsection{Vulnerable Library Characteristics}

As we mentioned before, we use the (library name, version) to build the connections between reused libraries of IoT firmware and vulnerabilities.
We analyze 10 popular reused libraries, including busybox, dnsmasq, and dropbear, which are widely used and have released hundreds of library versions, as listed in Table~\ref{tab:vld:rate}. 
We observe that these popular libraries have a high success rate (92\% on average) for extracting version information, indicating that their library headers usually store useful version information.

\begin{figure}[!t]
    \centering
      \includegraphics[width=2.5in]{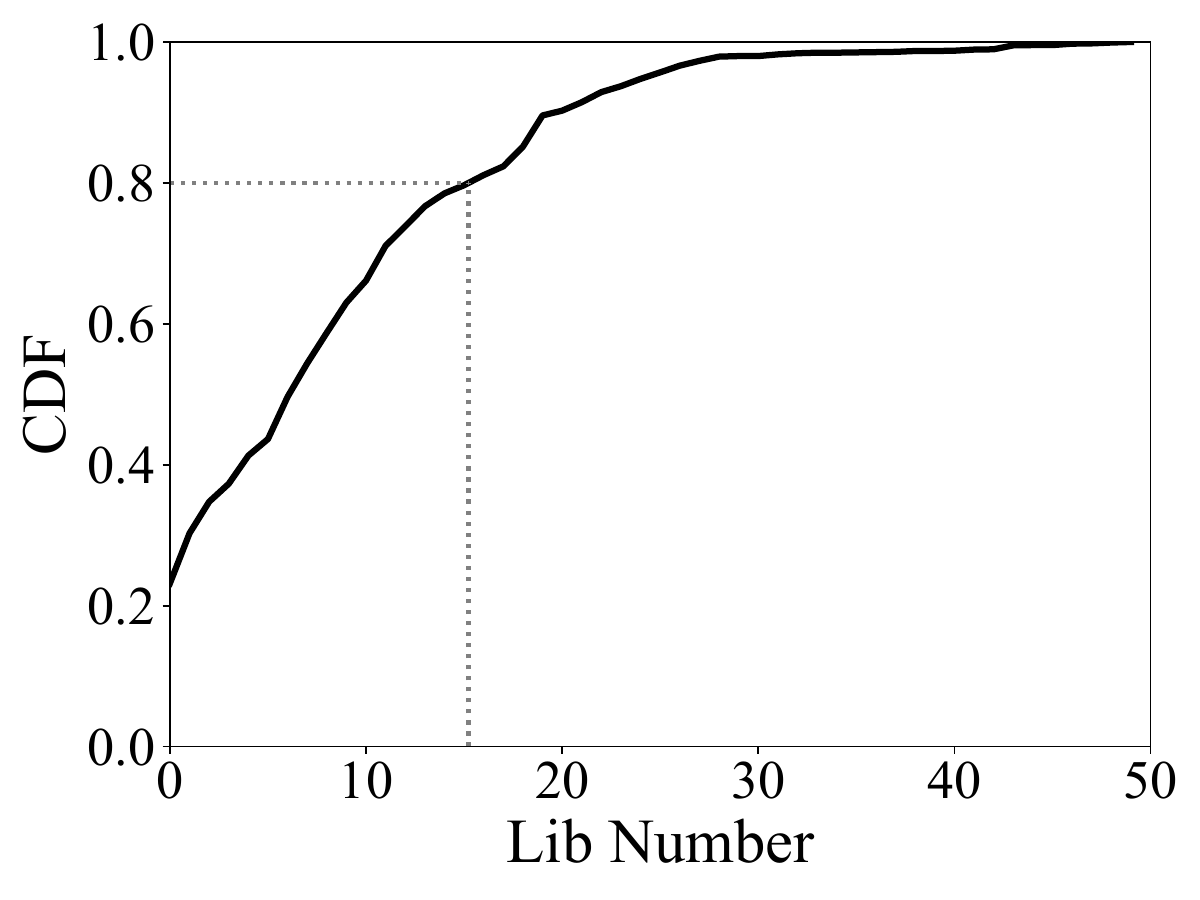}
             \caption{ The CDF of reused libraries for IoT firmware. The x-axis is the number of the vulnerable library version.}
       \label{fig:vld:syn}
\end{figure} 

\colorbox{mycolor}{\textit{Usage Analysis.}}
We present the distribution of the number of library versions reused by IoT firmware in Figure~\ref{fig:vld:syn}.
We find that 20\% of IoT firmware contains more than 15 vulnerable library versions, indicating a high risk of IoT devices.
By contrast, 23\% of IoT firmware in the dataset has 0 vulnerable library versions.
Those vulnerable libraries act as an attack surface for IoT devices, where attackers can exploit them to gain unauthorized access to devices, steal sensitive information, or launch attacks on other systems.

\begin{figure}[!t]
    \centering
      \includegraphics[width=2.5in]{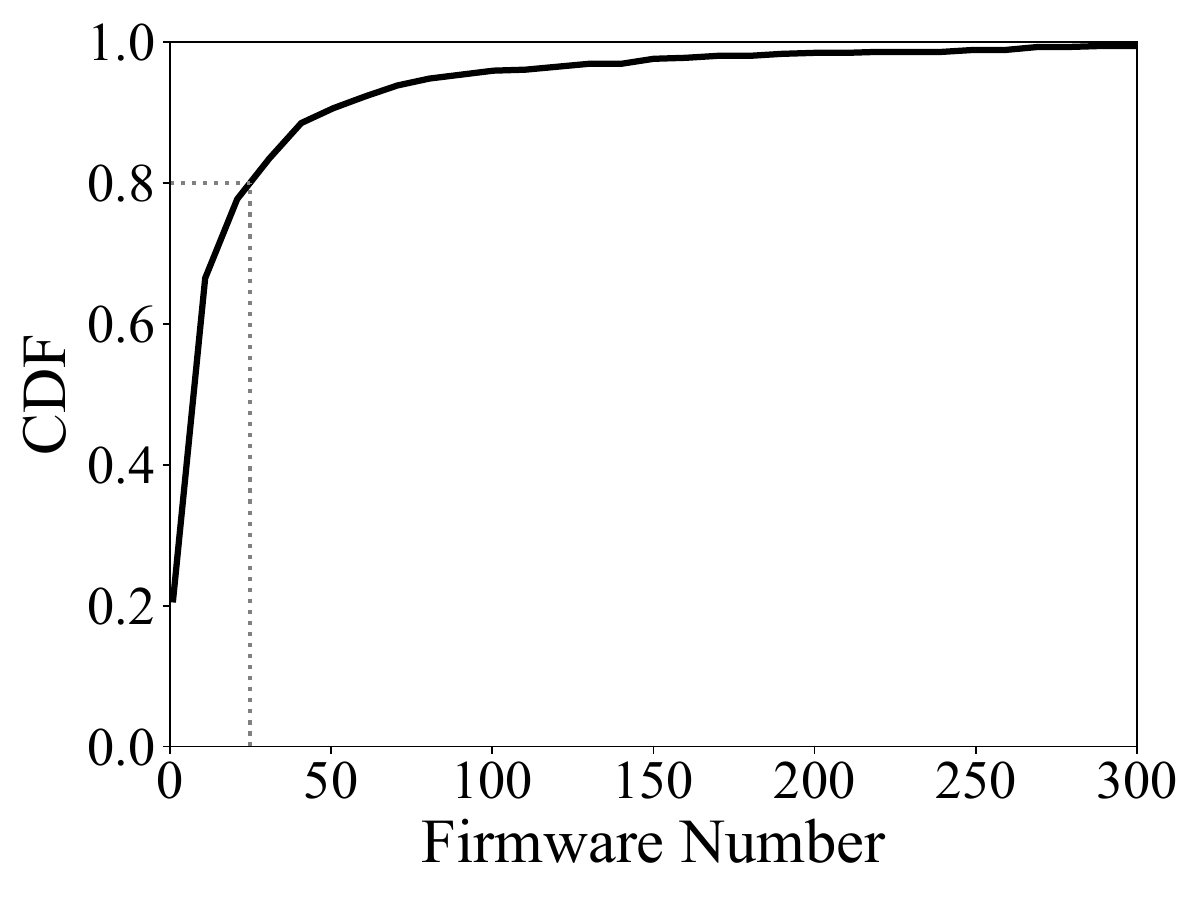}
         \caption{ The CDF of IoT firmware for vulnerable library versions.  The x-axis is the number of IoT firmware. The x-axis is the number of tvulnerable library. }
            \label{fig:lib:firmware}
\end{figure} 

We further present how many IoT firmware images reuse the same library version.
If a library version can be found in two IoT firmware, it is reused by two firmware.
Figure~\ref{fig:lib:firmware} is the CDF of firmware number for vulnerable library versions.
Results demonstrate that 20\% of library versions appear in more than 25 IoT firmware.  
The `bridge-utils' library version 1.0.6 is the most prevalent across 488 IoT firmware, followed by the `igmpproxy' library version 0.1 and `udpxy' library version 1.0.
The prevalence of vulnerable libraries highlights the need for developers to prioritize security in their development processes. 

\result{ It is common that IoT firmware consists of multiple vulnerable library versions and the same vulnerable library version is reused by a variety of IoT firmware.
}


\begin{table*}[!t] \small

    \caption{The overall vulnerabilities that are tightly bound with reused libraries in the IoT firmware.}
    \label{tab:vul:lib:firm}
    \centering
    \begin{tabular}{ c c c c | c c c c} 
    \toprule
          & \makecell[c]{Firm. \#Num.}  & \makecell[c]{Lib.  Num.}  &  \makecell[c]{CVE \#Num  (Distinct)} &       &   \makecell[c]{Firm. \#Num.}   & \makecell[c]{Lib.  Num.}  & \makecell[c]{CVE \#Num  (Distinct)}   \\
    \toprule
    360	& 16	& 172	& 1,990(50)	& ASUS	& 43	& 1,897	& 7,746(420)	\\
    Tomato-Shibby	& 312	& 3,519	& 30,356(1,447)	& DLink	& 70	& 961	& 7,170(239)	\\
    Foscam	& 3	& 7	& 51(4)	& Linksys	& 1	& 21	& 362(7)	\\
    Mercury	& 1	& 8	& 41(4)	& Mikrotik	& 5	& 10	& 12(3)	\\
    Netgear	& 123	& 1,271	& 15,956(488)	& Qnap	& 4	& 95	& 408(25)	\\
    Supermicro	& 27	& 155	& 2,148(64)	& Synology	& 29	& 665	& 6,368(140)	\\
    Tenda	& 124	& 581	& 8,325(281)	& Buffalo	& 3	& 37	& 592(13)	\\
    TP-Link	& 488	& 4,727	& 61,071(1,776)	& Trendnet	& 147	& 997	& 14,346(439)	\\
    Ubiquiti	& 369	& 3,134	& 65,245(1,691)	& Xerox	& 4	& 28	& 856(20)	\\
     \toprule
    \end{tabular}
\end{table*}

\section{Findings}

Starting from the initial set of our dataset in §II, we present the results of our study on vulnerable libraries in IoT firmware. 
Leveraging version information, we detected vulnerable libraries in 6,901 firmware images, resulting in a total of 18,285 library versions. 
Of these, 11,342 versions were found to contain vulnerabilities, with a total of 2,729 distinct CVEs.

Table~\ref{tab:vul:lib:firm} lists an overview of the vulnerability information for each IoT manufacturer, including the number of firmware images, the number of vulnerable libraries, and CVEs (distinct).
Our analysis revealed that the IoT manufacturer ``TP-Link'' had the largest number of CVEs (1,776) over their 448 IoT firmware, followed by ``Ubiquiti'' with 1,691 CVEs over 369 IoT firmware. 
This suggests that the more firmware a manufacturer releases, the higher the likelihood of vulnerabilities being present.
Our initial results indicate that the vulnerable versions are widely used in IoT firmware. 
On average, each firmware image had close to 2.3 CVEs resulting from the reused libraries, highlighting the significant impact that vulnerable libraries can have on IoT security. 
We hypothesize that this may be due to IoT manufacturers failing to update firmware for vulnerable versions of libraries.
To improve the security of their products, IoT manufacturers should prioritize timely updates of libraries in their firmware and adopt more robust software development practices.

To further explore the underlying risks for IoT firmware, we analyze the firmware images that contain vulnerable libraries.
Specifically, we aim to answer the following research questions.
\begin{itemize}
    \item \textbf{RQ1:} When IoT firmware reuses an outdated library version, would the manufacturer update to a newer version?
    \item \textbf{RQ2:} How long does an outdated/vulnerable library version persist in IoT firmware?
    \item \textbf{RQ3:} What is the impact and influence of those vulnerable versions of libraries in IoT firmware images? 
\end{itemize}

\subsection{RQ1: Reused Library Update}

\begin{figure}[!t]
    \centering
    \includegraphics[width=2.5in]{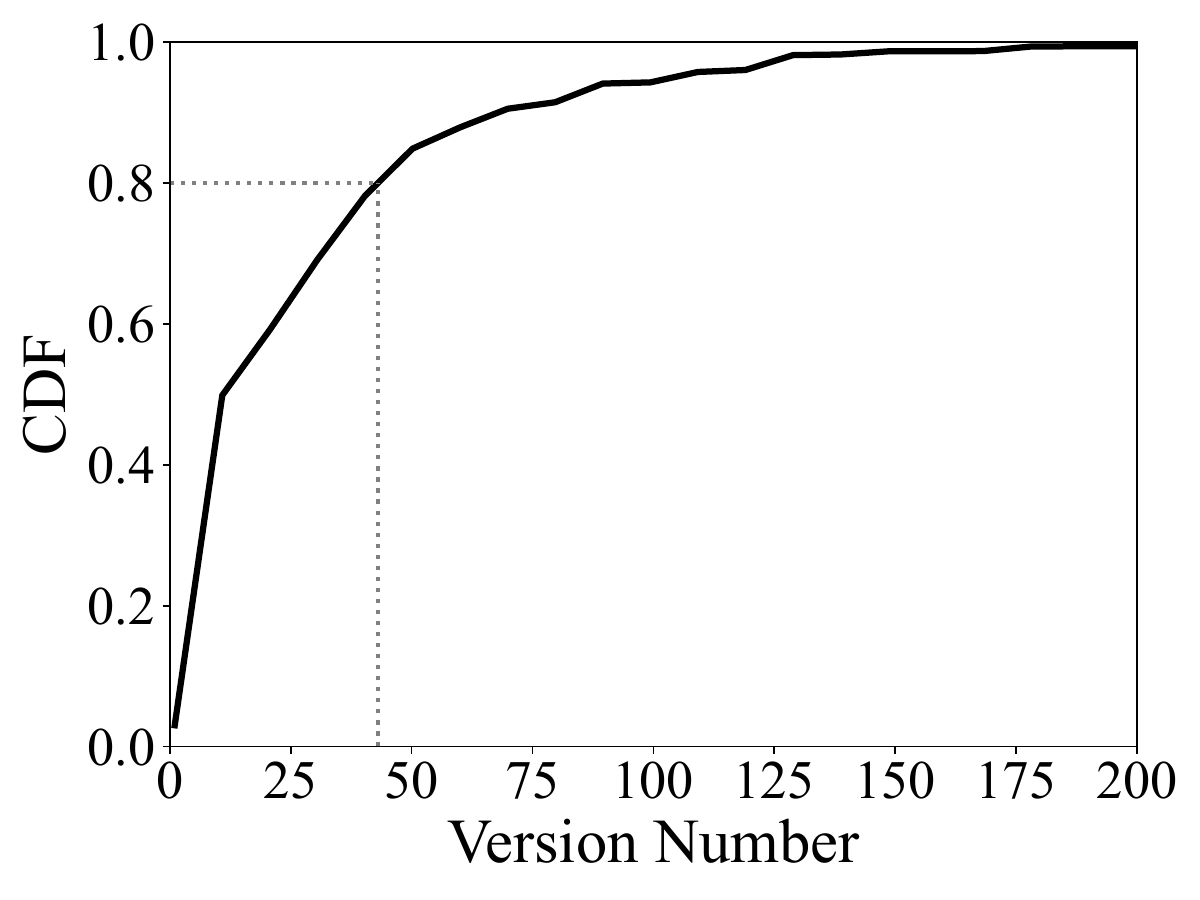}
    \caption{ The CDF distribution of outdated versions of software libraries in IoT firmware. 
    The x-axis is the number of differences between the versions.  }
    \label{fig:ver:outdated}
\end{figure}

\begin{figure}[!t]
    \centering
    \begin{tabular}{ c   c  } 
        \begin{minipage}[t]{0.47\linewidth}
            \includegraphics[width=1.0\linewidth]{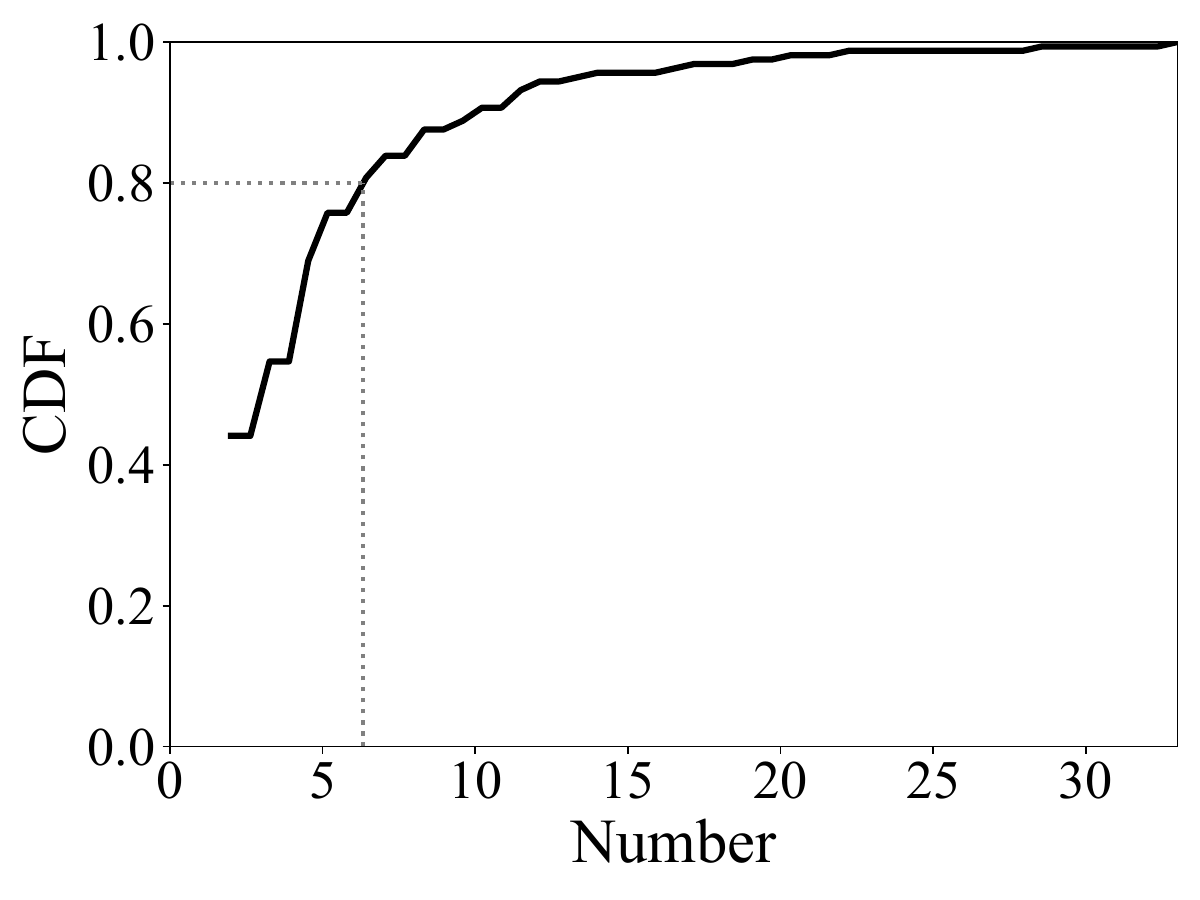}
            \caption{ The CDF of the number of reused library updates per IoT firmware update.  }
            \label{fig:ver:update}
        \end{minipage}
    &
    \begin{minipage}[t]{0.47\linewidth}
        \includegraphics[width=1.0\linewidth]{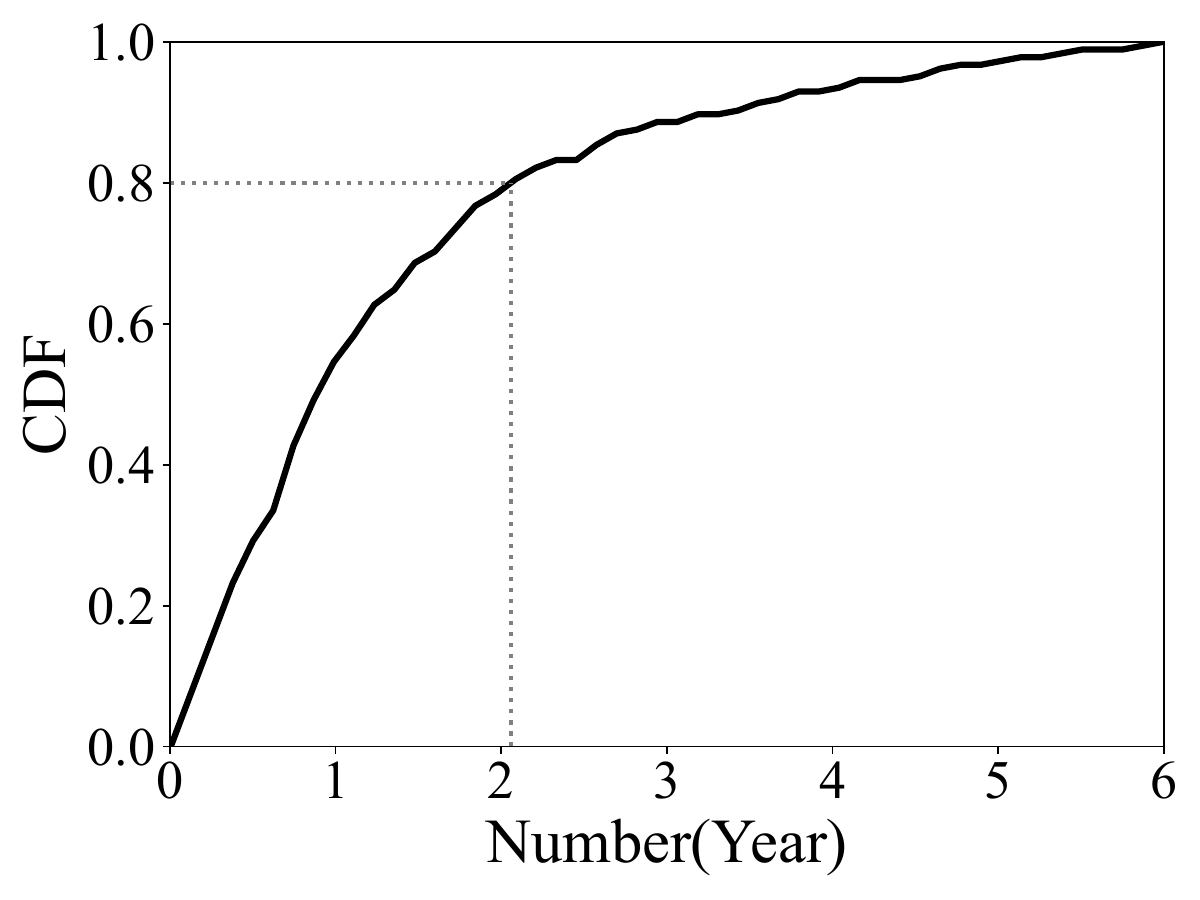}
        \caption{ The CDF of the time delay of reused library updates per IoT firmware.}
        \label{fig:update:time}
    \end{minipage}
    \end{tabular}
\end{figure}

When firmware reuses the outdated version of software libraries, IoT devices may cause underlying security risks.
We focused on identifying outdated library versions in IoT firmware and assessing the associated risks. 
Specifically, we considered the newest library version as the release version available on the official software website and compared it with the version used in the IoT firmware. 
If a library version is smaller than its newest version, the reused library is labeled as outdated.
We extract all outdated versions from IoT firmware.
To quantify the extent of this problem, we plotted the cumulative distribution function (CDF) of the version distance between the outdated version and the latest version for IoT firmware in Figure~\ref{fig:ver:outdated}. 
The figure shows that 80\% of the reused libraries in IoT firmware use an outdated version that is at least 50 version intervals away from the newest version.
This situation poses a significant risk to IoT firmware.
For instance, the ``busybox'' newest version is ``v1.36.0,'' and the extracted version is ``v1.9.2'' in IoT firmware ``DCS-7513\_REVB\_FIRMWARE\_v2.02.03'', where there are 150 versions of the intervals. 
We also investigated the reasons behind the prevalence of outdated versions in IoT firmware. Our inspection suggests that the manufacturers may not always maintain firmware and update it with the latest version of software libraries when it becomes available. 
This issue highlights the need for more proactive and regular firmware maintenance and updates to keep pace with the rapidly evolving software library landscape.

We further investigate whether manufacturers of IoT devices update outdated libraries in IoT firmware. 
To do this, we used a heuristic rule consisting of three steps: (1) we extract timestamps of all firmware images from a given vendor, (2) we determined which firmware images belonged to the same series of IoT devices based on their names, and (3) for each device series, we examined all reused libraries and their versions to determine if they had been updated.
The results of our investigation are concerning. In Figure~\ref{fig:ver:update}, we display the distribution of the number of reused library updates per IoT firmware. 
We found that 80\% of reused libraries were updated less than six times during the entire life cycle of the IoT device, and over 40\% of reused libraries were never updated. 
We also present the distribution of time delays for updating a newer library version of IoT firmware in Figure~\ref{fig:update:time}, which shows that 80\% reused libraries need more than two years to update their newer version.
These findings suggest that many outdated libraries may exist in IoT devices for relatively long periods, creating a significant risk for security vulnerabilities. Adversaries can exploit these vulnerabilities to access sensitive data, manipulate device functionality, or launch attacks on other systems.

\begin{figure}[!t]
    \centering
    \includegraphics[width=2.5in]{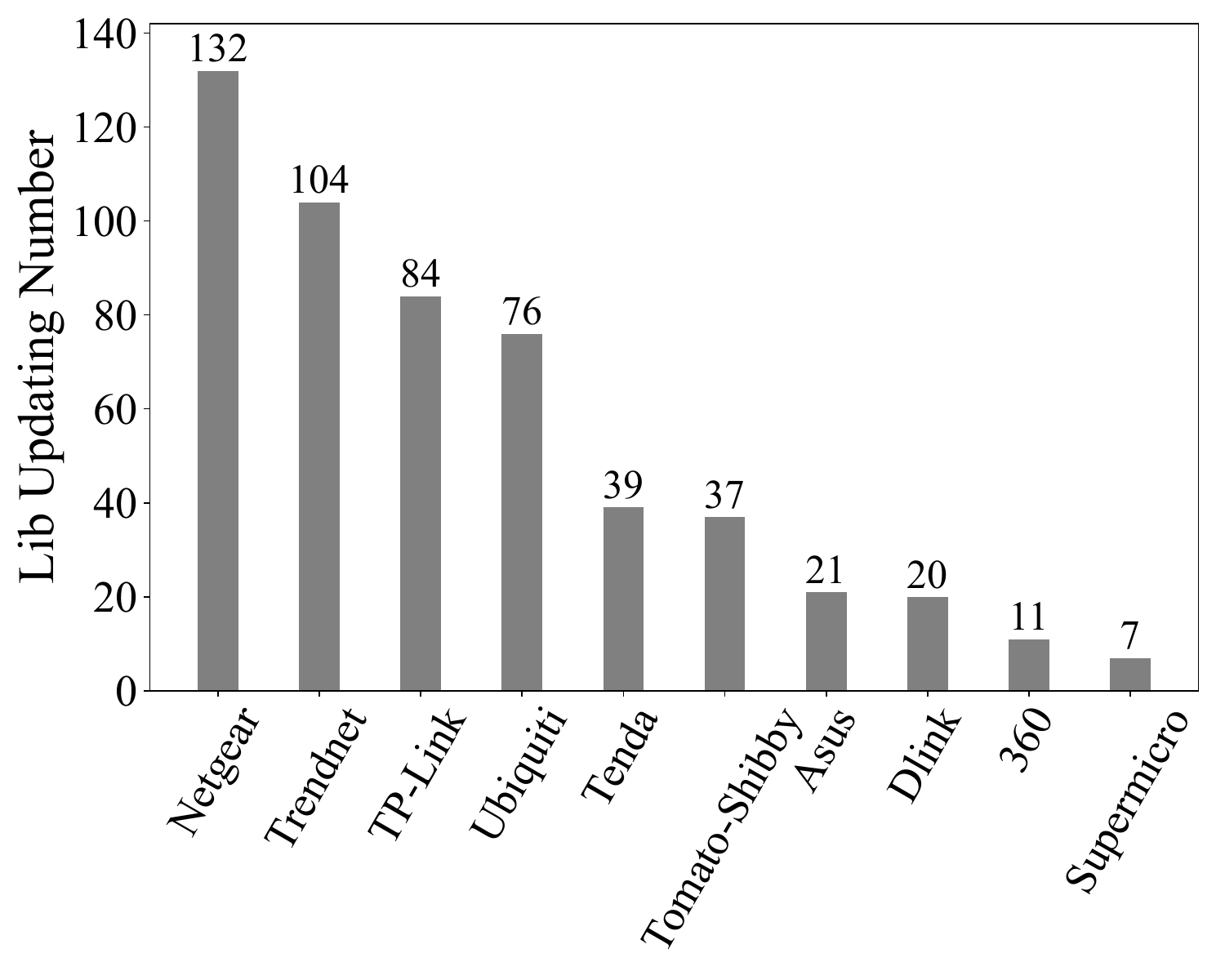}
    \caption{ The updating number of reused libraries across different manufacturers. }
    \label{fig:verndor:update}
\end{figure}

Updating to a newer library version ultimately depends on the developers of IoT manufacturers. Thus, we have categorized the updating numbers by the manufacturer in Figure~\ref{fig:verndor:update}. 
This graph shows that Netgear has the largest number of updates, followed by Trendnet and TP-Link. 
When a manufacturer releases a new IoT product or firmware, there is a higher likelihood of updating to a newer version of a reused library. However, we have also observed that some IoT vendors never update their library versions, usually because their firmware images are deprecated. 
It can be challenging to identify whether a firmware image is deprecated in our collected firmware. 
Though not significant, updating the library version in the IoT firmware is a rare operation for IoT manufacturers, indicating that attackers can still exploit the public vulnerabilities of IoT devices.
Therefore, it is crucial for manufacturers to prioritize regular updates to libraries in IoT firmware to ensure the security of their devices.

\finding{
    Overall, our findings showed that in 32.7\% of the cases, the manufacturer had updated the vulnerable library to a newer version. However, in 67.3\% of cases, the outdated/vulnerable library versions and vulnerabilities persisted in the firmware.}

\subsection{RQ2: Persistence Time}

When the software in IoT firmware is vulnerable, the adversaries can leverage their vulnerabilities to exploit IoT devices.
To better understand the risk of reused libraries, we analyzed the number of CVEs related to library versions in IoT firmware over time. 
The number of distinct CVEs in reused libraries has been increasing steadily, peaking at 399 in 2017. 
This trend is particularly concerning, as it indicates that the security of IoT devices is not improving over time. Despite efforts to improve the security of IoT devices, vulnerabilities related to reused libraries continue to persist and are even becoming more prevalent.
It's important to note that the number of distinct CVEs is not necessarily a perfect measure of the security of IoT devices. There may be many vulnerabilities that are not reported or that are reported but not assigned a CVE. Additionally, there may be multiple CVEs that are reported under the same library version.

\begin{figure}[!t]
    \centering
    \includegraphics[width=2.5in]{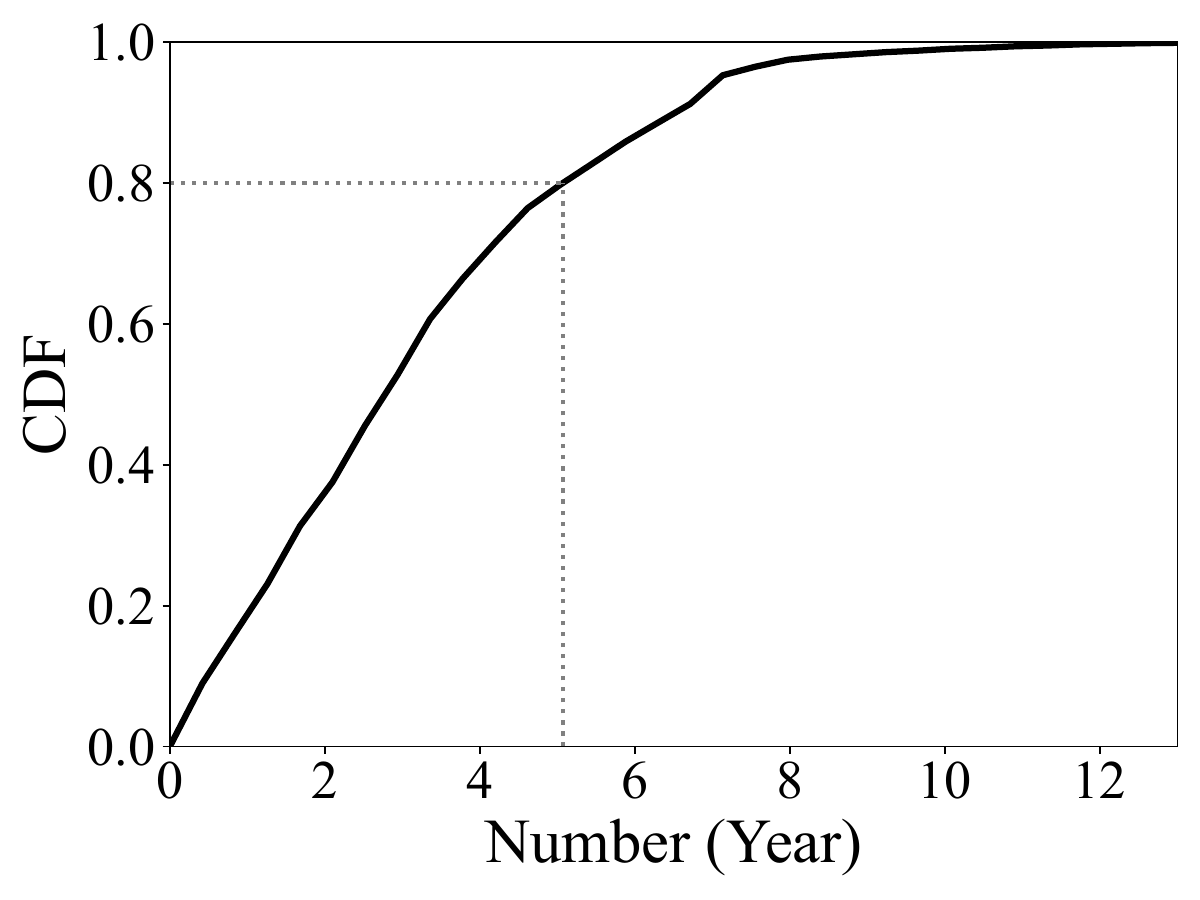}
    \caption{ The gap distribution between the time of vulnerability discovery and the IoT firmware publication time.  }
    \label{fig:software:diff}
\end{figure}

\begin{figure}[!t]
    \centering
    \includegraphics[width=2.5in]{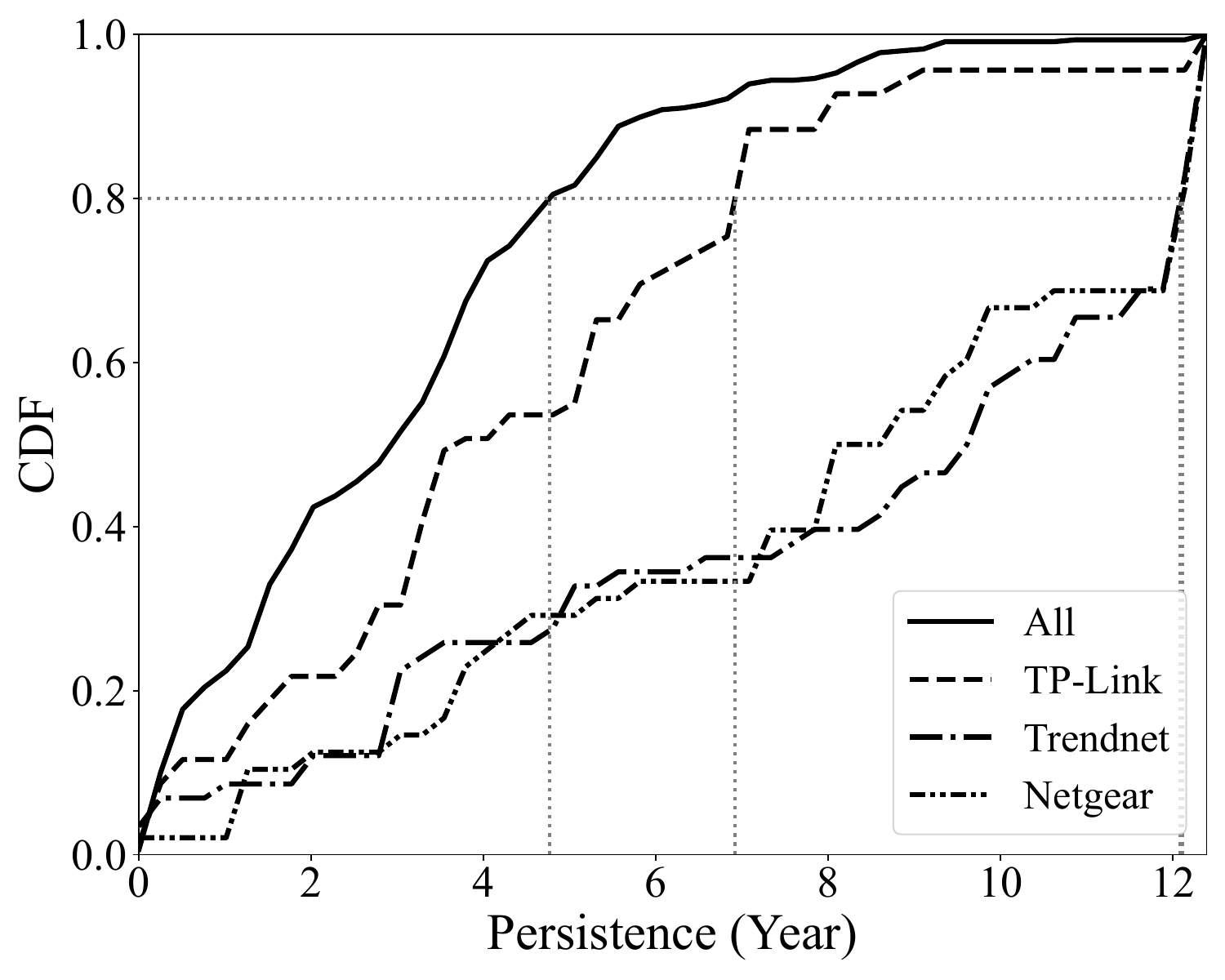}
        \caption{ The CDF distribution of the persistence time of IoT manufacturers.  }
        \label{fig:per:lib}
\end{figure}

Persistence delay is the time interval between the vulnerability discovered and the time when the IoT firmware image was released.
We use the persistence delay to present how long outdated/vulnerable versions exist in the IoT firmware.
Specifically, we obtain the discovered time of the vulnerabilities from NVD~\cite{nvd} and the released time of the IoT firmware from its metadata.
We use a heuristic rule: if the released time of the IoT firmware is earlier than the time of vulnerabilities discovered, this implies that those reused library versions have not been patched.

Figure~\ref{fig:software:diff} depicts the CDF distribution of the persistence delay, where the X-axis is the year unit, and the Y-axis is the cumulative probability. 
The largest persistence delay is up to 4,500 days (approximately 12 years). 
We can see that there are 80\% CVEs are at least 4 years behind the IoT firmware publication time.  
This is a significant amount of time for attackers to find and exploit those vulnerabilities. 
These findings suggest that IoT firmware manufacturers are slow in patching vulnerabilities, leading to the continued use of insecure library versions. 
The longer the delay, the higher the risk of potential attacks exploiting those vulnerabilities.
The fact that reused libraries with vulnerabilities persist for years in IoT firmware indicates a lack of focus on security from manufacturers.
It also highlights the challenges of securing IoT devices, as updating the firmware is not always straightforward, and manufacturers may not prioritize security patches.

As we mentioned before, the persistence delay of IoT firmware vulnerabilities is correlated with their IoT manufacturers.
We further present the distribution of the persistence time along with IoT manufacturers, as shown in Figure~\ref{fig:per:lib}, which describes three IoT vendors (TP-Links, Trendnet, and Netgear) for their persistence days of reused libraries. 
It highlights the fact that some manufacturers are more prone to fix vulnerabilities in their firmware than others, and this can have a significant impact on the security of their IoT devices.
For the manufacturer "TP-Link", 80\% of the vulnerabilities have at least a 6.5-year delay in fixing them.
On the other hand, the manufacturers ``Trendnet''  and ``Netgear'' have similar persistence delays for their device vulnerabilities, with a delay of almost 11 years. 
This suggests that these manufacturers may also not prioritize security as highly, and this could be a cause for concern for their users.

\finding{Our findings revealed that the average time taken for the manufacturer to update the vulnerable library was approximately 1.34 years. This suggests that manufacturers may not prioritize updating libraries in a timely manner, leaving firmware vulnerable to known attacks for extended periods.}

\subsection{RQ3: Vulnerability Impact}

\begin{figure}[!t]
    \centering
    \begin{tabular}{ c   c  }
        \begin{minipage}[t]{0.46\linewidth}
            \includegraphics[width=1.0\linewidth ]{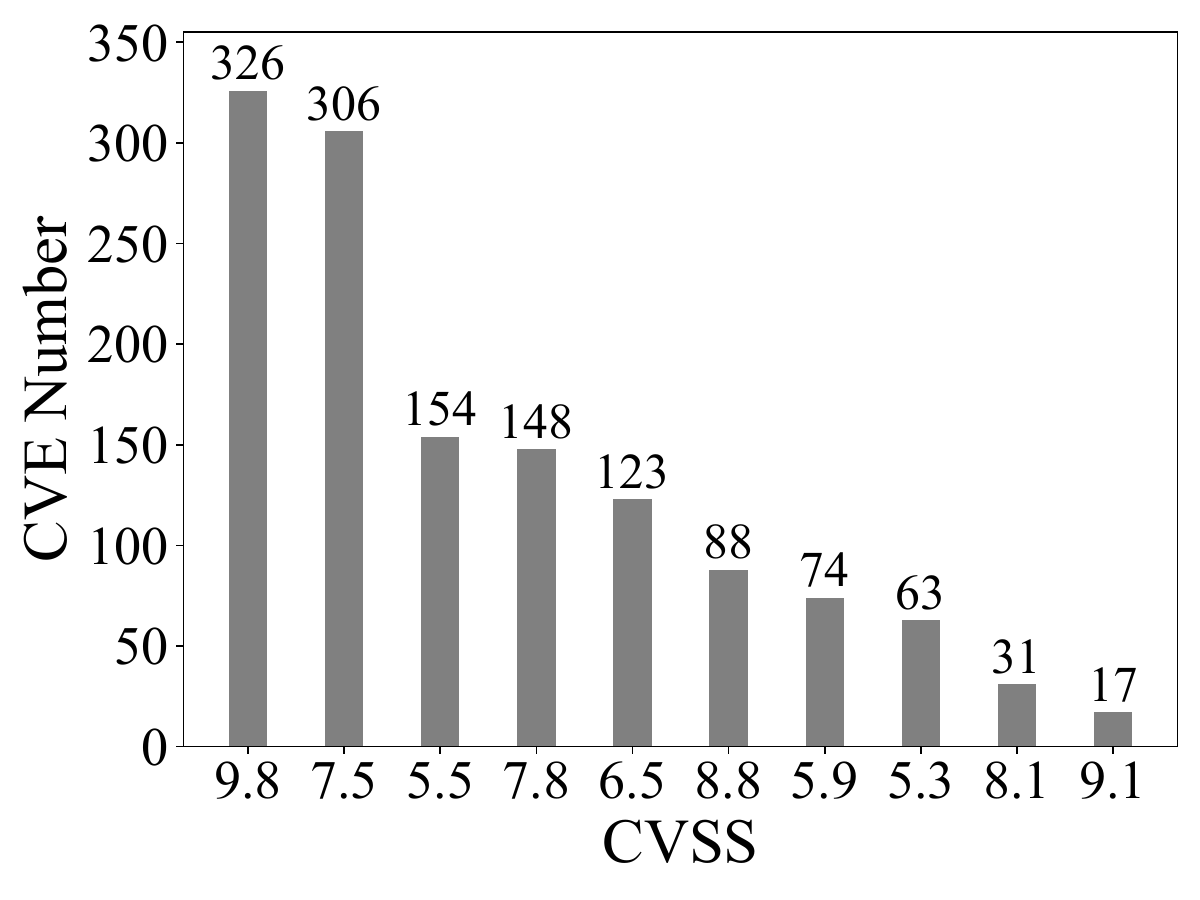}
    \caption{ The distribution of CVSS for vulnerabilities of reused libraries in the IoT firmware.  }
    \label{fig:cvss}
        \end{minipage}
    &
    \begin{minipage}[t]{0.46\linewidth}
        \includegraphics[width=1.0\linewidth ]{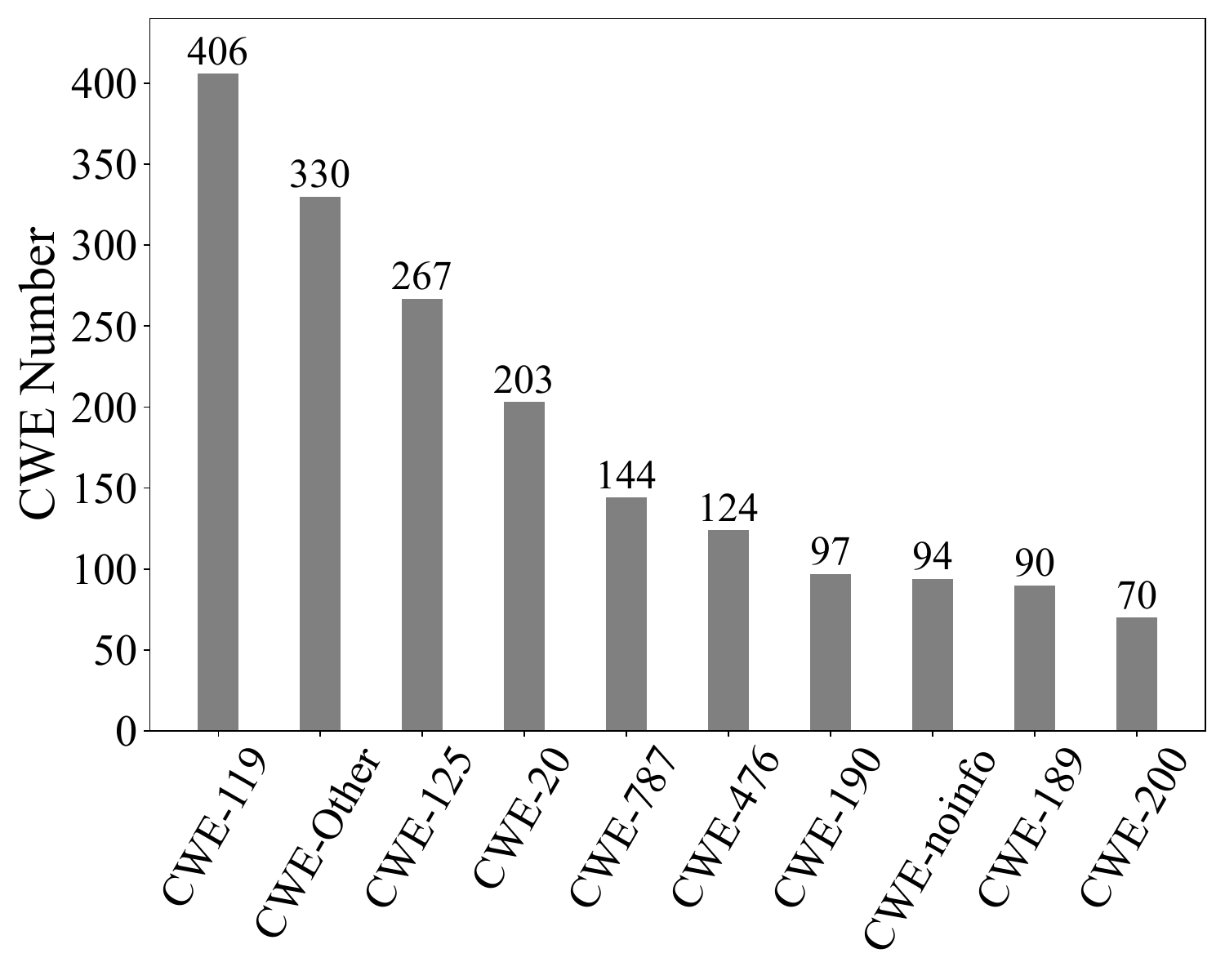}
    \caption{ The distribution of the vulnerability types for reused libraries in the IoT firmware. }
    \label{fig:software:cwe}
    \end{minipage}
    \end{tabular} 
\end{figure}

We use CVSS to present the vulnerability assessment, which comprises more than a dozen key characteristics, e.g., attack complexity, privilege required, user interaction, and confidentiality. For example, a CVSS score of 9.8 is considered critical, while a score of 7.5 is deemed high.
Figure~\ref{fig:cvss} depicts the distribution of CVSS for vulnerabilities of reused libraries in the IoT firmware. 
We find that many vulnerabilities have high CVSS values, where 298 vulnerabilities have a 9.8 score, followed by 244 vulnerabilities with a 7.5 score. 
Meanwhile, Figure~\ref{fig:software:cwe} shows the distribution of vulnerability types for these libraries, represented by the Common Weakness Enumeration (CWE) classification system.
Of all CWE types in the libraries, buffer overflow vulnerabilities (CWE-119) have the highest number of CVEs at 406. In addition, out-of-bounds read vulnerabilities (CWE-125) rank third with 267 CVEs.
Buffer overflow and out-of-bounds read vulnerabilities are common memory-related vulnerabilities that attackers can exploit to execute arbitrary code, crash a system, or leak sensitive data. 


In addition to CVSS scores, we assess their potential impact, particularly the scale of affected firmware number by the vulnerabilities of reused libraries. 
We leverage the Shodan~\cite{Shodan} to measure the popularity and usage of libraries with vulnerable versions on the Internet. 
Specifically, we identified the five most relevant vulnerabilities based on CVE-ID, library, version, and the number of detected firmware images by Shodan, as listed in Table~\ref{tab:shodan:lib}. 
Note that those firmware numbers may overlap and point to the same image. 
Our analysis shows that the combined usage of the most popular libraries, including tcpdump-4.9.2 (0.68 million), dnsmasq (0.44 million), dropbear (0.23 million), iptables (68 thousand), and busybox (3.4 thousand), suggests a significant impact of vulnerable libraries on IoT firmware.
Moreover, these vulnerabilities pose a significant threat as they can be exploited by attackers. 

\finding{Our findings suggest that vulnerabilities in these libraries have high CVSS scores, with a variety of risks, e.g., being exploited, unauthorized access to sensitive data, manipulation, or launching attacks.}

\begin{table}[!t]
    \caption{ Top-5 Vendors: the total number of libraries over four categories. }
    \label{tab:shodan:lib}
    \centering
    \begin{tabular}{c c c c  }
    \toprule
    CVE  & Library  & Version & Firmware number   \\
    \toprule
    CVE-2018-10105     & tcpdump    & 4.9.2 & 68,9053 \\
     CVE-2017-14495     & dnsmasq  & 2.71    & 417,335  \\
     CVE-2016-7408    & dropbear & 2011.54 & 421,048   \\
     CVE-2012-2663      & iptables & 1.4.4   & 239,603  \\
    CVE-2018-1000517   & busybox   & 1.21.1   & 3,465   \\  
    
     \toprule
    \end{tabular}
\end{table}

\section{Discussion \& Limitations}

In this section, we present three limitations of our {\tech}, including detection accuracy and coverage.

\textbf{Firmware Limitation}.   
Downloading the firmware is only sometimes available; even when possible, the firmware can have embedded proprietary file formats that are not easily extracted. 
Some popular IoT manufacturers do not provide end users with download links, making obtaining firmware images even harder. 
In the future, we can explore alternative methods, such as sniffing OTA (Over the air) and dumping the device's memory to collect more firmware images.

\textbf{Library Extraction Limitation}. 
Uncommon filesystems and compressed algorithms can hinder converting firmware into the filesystem to find relevant libraries. 
Existing tools like Binwalk used in {\tech} may not be able to extract the filesystem for IoT firmware with uncommon formats. To address this, we can find and integrate vendor-specific supportable formats into {\tech} to obtain more filesystems. 
Furthermore, the coverage of the candidate set of reused software libraries can be limited, and continuously constructing the dataset of reused libraries for IoT firmware can be an ongoing challenge.

\textbf{Version Detection Limitation}. 
{\tech} relies on version information to determine vulnerable software libraries in IoT firmware, but version information is not always available. There are three causes affecting the {\tech}'s performance.
(1) Many CVEs have no version information in our dataset, with nearly 2,469 vulnerabilities and 2,436 distinct CVEs. For instance, the CVE-2001-0965 has no version of the software `asterisk 13.3.2'.
(2) Library versions in IoT firmware are inconsistent with the version from CVE.  
Version errors and inconsistencies would impact the performance of our analysis, which needs to be further calibrated. 
However, calibrating version information requires knowledge of the background of IoT firmware and vulnerability.

\section{Related work}

\textbf{Firmware}. 
Several prior works detected vulnerabilities in the firmware via the static~\cite{helmke2023towards, feng2023aim, wuyour} and dynamic analysis~\cite{zhang2019cryptorex}. 
\citet{davidson2013fie} proposed the symbolic execution to cover possible execution paths for finding vulnerabilities in firmware, specifically on the MSP430 device. 
\citet{costin2014large, costin2016automated} are the first to conduct a large-scale analysis of firmware, which searches for specific bugs with apparent features but cannot handle more general cases.
\citet{shoshitaishvili2015firmalice} performed extensive static analysis on small slices of code and detected authorization bypass vulnerabilities in binary firmware. 
\citet{chen2016towards} collected numerous firmware images and performed automatic vulnerability verification over those firmware files.
\citet{feng2016scalable, feng2020p2im} leveraged the high-level numeric feature comparison to find vulnerabilities for 
firmware images. 
\citet{camurati2018inception} proposed the symbolic execution to test software security via an intermediate representation of an embedded system.
\citet{redini2020karonte} leveraged static analysis techniques to perform multi-binary taint analysis for embedded systems.
In contrast, our paper focused on the reused libraries in IoT firmware; more specifically, we leverage the syntax to search vulnerabilities caused by the firmware libraries.

\textbf{Binary Similarity}. 
Typically, the binary similarity is to compare two binary codes to identify their similarities and differences, e.g., basic blocks, functions, or whole programs. 
\citet{xu2017neural} proposed the neural network to learn the similarity classification between two binary codes. 
Similarly, \citet{david2018firmup} used binary code similarity to find firmware bugs in a large repository of target pieces of binary codes. 
\citet{yu2020order} proposed the pre-trained embedding and semantic neural network to compute the similarity of two binary codes.  
\citet{yu2020codecmr} further constructed a hybrid model by combining GNN and CNN to improve the performance of binary similarity.
\citet{ren2021unleashing} investigated the impact of non-standard optimizations while compiling source codes into the binary code, which brought adversarial examples for binary similarity.  
\citet{zhao2022large, zhao2023one} leveraged syntactical and control-flow graph features to calculate the binary similarity and provided a public/available dataset for the IoT firmware. 
\citet{marcelli2022machine} compared the state-of-the-art binary similarity approaches based on the graph-based embedding models and graph neural networks.
The binary similarity approach suffers from several long-standing challenges in the general case, e.g.,  distinguishing code from data and indirect control flow resolution. 
In this paper, we use syntax information (\textit{name, version}) designed for combining IoT firmware libraries and vulnerabilities at a large scale.

\textbf{Library Detection}.
There are many prior works to detect third-party libraries in Android or embedded systems~\cite{obaidat2024daedalus, zhao2023uvscan}. 
\citet{backes2016reliable} detected third-party libraries in Android applications based on abstracted package trees and method signatures, providing a database to the public. 
\citet{lauinger2018thou} investigated the inclusion of libraries with known vulnerabilities in websites and found 37\% of the websites at least have one vulnerable library.
ATVhunter~\cite{zhan2021atvhunter} proposed that several parallel root packages as interdependent parts constitute a third-party library and proposed version-based detection for vulnerabilities in Android. 
\citet{zhan2020automated, zhan2021research} conducted a comprehensive comparison of 11 detection tools as a benchmark for third-party libraries in Android.
\citet{yang2022modx} leveraged program modularization techniques to decompose the program into functional features for detecting third-party libraries.
\citet{zhang2021capture} proposed an in-hub security manager to capture software library patches for IoT firmware. 
In contrast to Android or the Web, we are to provide a systematic analysis of third-party libraries in IoT firmware and find vulnerabilities by library versions.

\section{Conclusion}

With the increasing number of IoT firmware and software packages, there is a growing concern over the security risks caused by insecure software libraries in IoT systems.
In this paper, we conduct a large-scale empirical study on 6,901 IoT firmware images, evaluating over 11,342 vulnerable versions across a diverse set of 349 software projects.
The investigation centered around a dataset we collected that merges IoT firmware from manufacturers' websites, vulnerability entries from the NVD, and affected software in the filesystem.
Using this dataset, we further systematically analyze firmware, software libraries, and related vulnerabilities.
Our findings are that (1) the IoT manufacturer rarely updates a new version for an outdated version/vulnerable of the library, (2) outdated/vulnerable versions and vulnerabilities have a long persistence time (several years) in the IoT firmware, and (3) vulnerabilities of software libraries have posed server threats to widespread IoT devices.

\bibliographystyle{ACM-Reference-Format}
\bibliography{ref}
\clearpage
\appendix

\section*{Online Source}

\begin{table}[htb] \small
    \centering
    \caption{ IoT firmware: device vendor and device type.}
    \label{tab:download:list}
    \begin{tabular}{ c c }
      \toprule
  Vendor   &  Device Type  \\
      \midrule
  360           & Router                           \\
  asus          & Router                           \\
  buffalo       & Access, Router                    \\
  dlink         & Router, Switch, Access, Camera      \\
  foscam        & Camera                           \\
  linksys       & Switch, Router                    \\
  mercury       & Camera, Router, Access, Switch      \\
  microstrain   & Sensors                          \\
  mikrotik      & Router, Switch                    \\
  netgear       & Router, Firewall, Switch, Extender  \\
  qnap          & OS, Storage, Switch, Router         \\
  se            & Automation                       \\
  supermicro    & BIOS, BMC, Bundle                  \\
  synology      & OS                               \\
  tenda         & Access, Adapter, Camera, Router     \\
  ti            & OS                               \\
  tomato-shibby & Router                           \\
  tp-link       & Router, Access, Adapter, Camera     \\
  trendnet      & Access, Adapter, Camera, Controller \\
  ubiquiti      & Router, Unifi, Uvc, Ufiber          \\
  xerox         & Printer                         \\
   \bottomrule
  \end{tabular}
\end{table}

\begin{table}[hbt] \small
	\caption{ Firmware Metadata}
	\label{tab:mata}
	\centering
	\begin{tabular}{c c}
		\toprule
		\multicolumn{2}{c}{An example of IoT firmware metadata.}	\\
		\midrule
		Firmware Name & DCS-6517\_REVB\_FIRMWARE\_v2.01.00 \\
		Manufacturer & D-Link  \\
		Device type & Camera \\
        Product  & DCS-6517 \\
        Version & V2.01.00 \\
        Publish time & 2020-07-09 \\
        URL & \makecell[l]{ftp://FTP2.DLINK.COM/PRODUCTS/DCS-6517/\\REVB/DCS-6517\_REVB\_FIRMWARE\_v2.01.00.zip}\\
		checksum & 8773593588fcb789c88d8275b49d7d7f \\
		\bottomrule
	\end{tabular}
\end{table}

\end{sloppypar}
\end{document}